\documentclass[%
 reprint,
 superscriptaddress,
 amsmath,amssymb,
 aps, prx,
]{revtex4-2}

\usepackage{graphicx}
\usepackage{dcolumn}
\usepackage{bm}
\usepackage{xcolor}
\usepackage{booktabs}
\usepackage{hyperref}

\begin{document}

\author{Niclas Krupp}
\email{niclas.krupp@pci.uni-heidelberg.de}
\affiliation{%
 Theoretische Chemie, Physikalisch-Chemisches Institut, Universität Heidelberg, INF 229, 69120 Heidelberg, Germany 
}%
\author{Gerrit Groenhof }
\affiliation{%
Nanoscience Center and Department of Chemistry, University of Jyväskylä, P.O. Box 35, Jyväskylä 40014, Finland
}%
\author{Oriol Vendrell}
\email{oriol.vendrell@uni-heidelberg.de}
\affiliation{%
 Theoretische Chemie, Physikalisch-Chemisches Institut, Universität Heidelberg, INF 229, 69120 Heidelberg, Germany 
}%

\title
  {Quantum dynamics simulation of exciton-polariton transport}


\begin{abstract}
    Strong coupling between excitons and confined modes of light presents a promising pathway to tunable and enhanced energy transport in organic materials. By forming hybrid light-matter quasiparticles, exciton-polaritons, electronic excitations can traverse long distances at high velocities through ballistic flow. However, transport behavior of exciton-polaritons varies strongly across experiments, spanning both diffusive and ballistic transport regimes. Which properties of the material and light-modes govern the transport behavior of polaritons remains an open question.
    Through full-quantum dynamical simulations we reveal a strong dependence of polariton transport on vibronic interactions within molecules in both ideal and lossy cavities. Specifically, we show that intramolecular vibrations mediate relaxation processes that alter polariton composition, lifetime and velocity on ultrafast timescales. Analysis of the propagating wavepacket in position and momentum space provides mechanistic insight into the robustness of ballistic flow of exciton-polaritons
    found experimentally under cryogenic conditions.

\end{abstract}
\maketitle
\section{Introduction}
Since the seminal observation of Rabi splitting in a semiconductor microcavity in 1992 \cite{weisbuch_observation_1992}, the field of exciton-polaritons (EPs) -- hybrid light-matter states involving electronic material excitations and quantized modes of light -- 
has been rapidly evolving in both theory and experiment. Nowadays, experimental platforms which typically consist of a Fabry-Pérot microcavity filled with inorganic or organic emitters can routinely achieve strong coupling between material and photonic subsystems \cite{bhuyan_rise_2023,khazanov_embrace_2023,jiang_excitonpolaritons_2022}. Consequently, a growing body of research is now dedicated to the characterization and application of the unique properties of EPs, bridging condensed matter physics, material science and chemistry. EPs have low effective mass and exhibit adjustable dispersion relations that differ from those of the bare material. They can also be selectively optically excited, which enables the development of highly tunable platforms for exciton condensation \cite{byrnes_excitonpolariton_2014, su_observation_2020, daskalakis_spatial_2015, plumhof_room-temperature_2014, de_room-temperature_2023, zasedatelev_room-temperature_2019,laitz2023uncovering} and transport processes \cite{xu_ultrafast_2023, balasubrahmaniyam_enhanced_2023, lerario2017high, myers2018polariton, rozenman2018long, sokolovskii_multi-scale_2023}. Hybridization of excitons with cavity photons might overcome notoriously inefficient and diffusive exciton transport in organic materials, achieving ballistic energy flow in polaritonic devices \cite{sanvitto_road_2016}. Thus, polariton transport presents a strategy to improve and steer exciton mobility in semiconductors, which is of high relevance for organic photovoltaics \cite{rafique_fundamentals_2018}. Indeed, experimental studies on EP transport could show propagation of EPs over long distances, and over time spans which exceed the bare cavity photon lifetime \cite{xu_ultrafast_2023,balasubrahmaniyam_enhanced_2023,lerario2017high,myers2018polariton}.

However, the mechanism and efficiency of polariton transport are still unclear: ballistic EP flow has been observed in inorganic semiconductor polaritons at cryogenic temperatures \cite{freixanet_-plane_2000,myers2018polariton}, and for organic Bloch surface wave polaritons \cite{lerario2017high}. In these cases, polaritons move at the expected group velocity obtained from the gradient of the dispersion curve. In contrast, other studies in organic J-aggregate microcavities measured diffusive EP transport at reduced velocities below the expected group velocity \cite{rozenman2018long}.
Current state-of-the-art ultrafast microscopy experiments on inorganic microcavity polaritons and organic Bloch surface wave polaritons found a transition from ballistic to diffusive transport which can be tuned via the exciton content of EP \cite{xu_ultrafast_2023,balasubrahmaniyam_enhanced_2023}. Here, ballistic transport at high group velocities is maintained for up to 50\% exciton content, while more excitonic EPs propagate diffusively at lower-than-expected velocities. In the low-temperature regime, steady-state measurements \cite{freixanet_-plane_2000,myers2018polariton} and recent time-resolved experiments \cite{xu_ultrafast_2023} both have found ballistic flow and no group velocity reduction, even at exciton contents as high as 88\% \cite{xu_ultrafast_2023}.

Organic condensed-phase microcavities exhibit static and dynamic disorder. Static disorder, on the one hand, is connected with the material-dependent inhomogeneous broadening of the excitonic resonance.  Dynamic disorder, on the other hand, results from molecular nuclear dynamics and results in time-dependent fluctuations in electronic transition energies and cavity-molecule interaction. Nuclear dynamics can either occur thermally-activated in the electronic ground state, or in the excited states due to different topologies of ground and excited state potential energy surfaces at the Franck-Condon point, i.e., due to vibronic coupling. Both types of disorder lead to elastic and inelastic scattering processes of EPs, such as vibration-assisted scattering and radiative pumping, which are ubiquitous in semiconductor cavity systems \cite{coles2011vibrationally,agranovich_cavity_2003,coles2013imaging,somaschi2011ultrafast,luttgens2020population}. Yet, their impact on EP dynamics is still debated in the field.
Focusing on EP transport, dynamic and static disorder effects -- in particular EP-phonon and EP-defect scattering -- are speculated to play a role in determining the ballistic to diffusive crossover, group velocity renormalization, and the feeding of population from the exciton reservoir back to the coherent polaritonic manifold \cite{virgili_ultrafast_2011,xu_ultrafast_2023,sokolovskii_multi-scale_2023,pandya_microcavity-like_2021}. 

From a theory perspective, studies based on 
semiclassical Ehrenfest-type dynamics \cite{sokolovskii_multi-scale_2023,sokolovskii_photochemical_2024,tichauer_tuning_2023} could gain first insights into effects of cavity losses and reversible vibration-driven population transfer to dark states on EP transport. However, simulations entirely based on quantum mechanics and including molecular vibrations and dispersive cavity modes have not been attempted yet, but are clearly needed for a thorough mechanistic understanding of polaritonic transport processes. Here, we report full-quantum dynamical simulations of exciton transport in molecular arrays coupled to perfect and lossy Fabry-Pérot cavities. Employing the highly compact multi-layer multiconfiguration time-dependent Hartree (ML-MCTDH) wavefunction ansatz  \cite{wan_03_1289,man_08_164116,vendrell_multilayer_2011,vendrell_coherent_2018} enables us to account for the multi-mode nature of the cavity, and more than 200 molecules with vibrational degree of freedom. The molecular ensemble-cavity wavefunction is then propagated according to the ML-MCTDH equations of motion which are obtained from the time-dependent variational principle \cite{wan_03_1289}. With this efficient and accurate simulation protocol at hand, we can closely examine the impact of dynamic disorder, radiative decay and laser excitation parameters on EP transport in an ideal organic crystal within a fully quantum-dynamical framework. 

Experiments under cryogenic conditions are closely comparable to  our quantum-dynamics simulations at 0~K. Our results thus offer detailed insight into preserved ballistic transport of polaritons in the low-temperature regime. Importantly, our full quantum-dynamical simulations support the existence of ballistic transport even for highly excitonic initial states, while uncovering the role of vibronic coupling at low temperatures.
Molecular vibrations which couple to electronic transitions can improve EP transport by mediating relaxation of highly excitonic wavepacket components toward increasingly photonic states with higher group velocity.
By considering experimentally relevant excitation schemes with explicit laser pulses, we identify regimes in which this vibronic-enhancement mechanism is operative. Moreover, we find that this effect can revert and lead to a slow-down of EP transport, depending on the targeted polaritonic branch. 
Finally, our simulations suggest that resonant excitation of EPs on the lower polariton branch close to 50\% exciton content, provides a favorable trade-off between high propagation velocities and prolonged polariton lifetime in lossy cavities with realistic radiative lifetimes.

\section{\label{sec:level1}Theory}
\begin{figure*}
    \centering
    \includegraphics[width=0.9\textwidth]{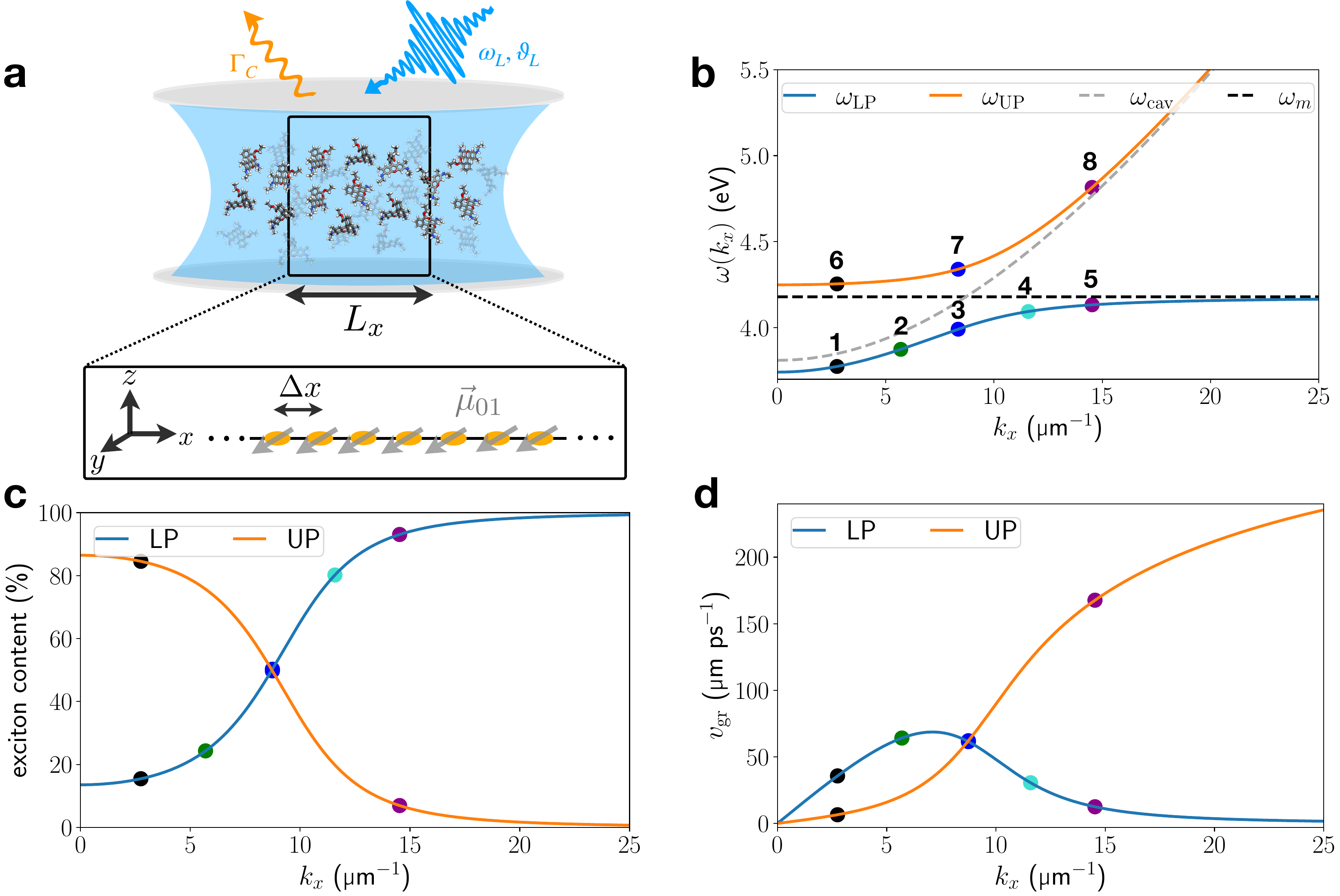}
    \caption{(a) Schematic of an organic microcavity filled with an ordered molecular ensemble in which all molecular TDMs $\vec \mu_{01}$ are aligned with the cavity polarization direction. (b) Polariton dispersion curve $\omega_{\mathrm{LP/UP}}(k_x)$ with bare molecular $\omega_m$ and bare cavity $\omega_{\mathrm{cav}}$ dispersion. (c) Corresponding exciton content and (d) group velocities of lower and upper polariton branches. Colored circles in (b)-(d) mark the polaritonic states targeted through resonant excitation by an external laser pulse.}
    \label{fig:Figure_1}
\end{figure*}

In the following, we give a brief overview over the theory of exciton-polaritons and  the details of our organic microcavity model. We consider a one-dimensional array of $N$ identical molecules inside the polarization-plane of Fabry-Pérot (FP) cavity, described by the Hamiltonian
\begin{align}
    \hat H &= \hat H_{\mathrm{cav}} + \hat H_{\mathrm{mol}} + \hat H_{\mathrm{cav-mol}}.
\end{align}
The electromagnetic field is assumed to be confined in the $z$-direction, and molecules are placed equidistantly along the $x$-axis as depicted in Fig.~\ref{fig:Figure_1}a.
Periodic boundary conditions along $x$ are assumed. Following recent molecular dynamics studies on polariton transport \cite{sokolovskii_multi-scale_2023,tichauer_tuning_2023}, we further restrict the FP cavity model to a single polarization direction, $\vec\epsilon_y$ and a single propagation direction ($k_x\geq 0$). Thus, the Hamiltonian of the cavity field reads
\begin{align}
\hat H_{\mathrm{cav}} & = \sum_{k_x} \omega_{\mathrm{cav}}(k_x)\hat a^{\dagger}_{k_x}\hat a_{k_x}  \label{eq:hcav}
\end{align}
with bosonic creation (annihilation) operators $\hat a^{\dagger}_{k_x}$ ($\hat a_{k_x}$) which create (annihilate) a cavity photon with in-plane momentum $k_x$ and energy $\omega_{\mathrm{cav}}(k_x)$, given by the cavity dispersion relation $\omega_{\mathrm{cav}}(k_x) = \sqrt{\omega_0^2 + c^2k_x^2/n^2}$, where $\hbar\omega_0$ is the cavity mode energy at normal incidence ($k_x=0$), $c$ the speed of light and $n$ the refractive index of the medium. Importantly, an appropriate discretization and truncation of in-plane momentum values $k_x$ is required in practice. Note that imposing periodic boundary conditions naturally discretizes in-plane momenta according to $k_{x,p}=2\pi p/L_x$ where $L_x = N\Delta x$ and $p = 0,1,\dots,M-1$ \cite{Michetti2005}. In our simulations we account for $N=256$ molecules and include $M=256$ photonic modes. Convergence with respect to these numbers has been checked in advance (Fig.~S1).

Each molecule is described by a Holstein-type model Hamiltonian consisting of electronic ground (S$_0$) and excited state (S$_1$) as well as one vibrational degree of freedom $Q_j$. For both electronic states, harmonic potential energy curves (PEC) along $Q_j$ with identical vibrational frequency $\omega_{\mathrm{vib}}$ are employed. Varying horizontal displacements of the excited state PEC with respect to the ground state PEC are investigated, which enter the molecular Hamiltonian as linear vibronic coupling constants $\kappa$. Consequently, the molecular Hamiltonian reads
\begin{align}
\hat H_{\mathrm{mol}} & = \sum_{j=0}^{N-1}  \omega_m \hat \sigma_j^{\dagger}\hat\sigma_j + \frac{\omega_{\mathrm{vib}}}{2}\left(-\frac{\partial^2 }{\partial Q_j^2} + Q_j^2\right) + \kappa Q_j \hat \sigma_j^{\dagger}\hat\sigma_j,  \label{eq:hmol}
\end{align}
where $\hat \sigma_j^{\dagger}$ ($\hat \sigma_j$) create (annihilate) an electronic excitation (exciton) at the $j$-th molecule in the S$_1$ state with vertical electronic excitation energy $\omega_m$ at the Franck-Condon point. The vibronic coupling constants $\kappa$ considered here (0\,meV, 50\,meV, 80\,meV) correspond to Huang-Rhys factors of $S=0.0,0.15,0.58$, respectively, reflecting the vibronic couplings in typical organic molecules such as pyrazine, pyrene, or Rhodamine derivatives \cite{worth1996effect,terry2023theory,barclay2022characterizing}. Furthermore, since a sufficiently large distance $\Delta x$ is assumed, intermolecular interactions (for instance, exciton-exciton coupling) can be neglected. This prohibits any intermolecular exciton transport in the absence of coupling to the cavity.

The interaction between the cavity field and molecular excitation is described without invoking the rotating wave approximation \cite{cwik_excitonic_2016,sokolovskii_multi-scale_2023} by
\begin{eqnarray}
\hat H_{\mathrm{cav-mol}} = &&\sum_{j=0}^{N-1}\sum_{k_x} g_{j}(k_x)\left(\hat a^{\dagger}_{k_x}\mathrm{e}^{-ik_x x_j}+\hat a_{k_x}\mathrm{e}^{ik_x x_j}\right)\nonumber\\
&&\times(\hat\sigma_j^{\dagger}+\hat\sigma_j). \label{eq:hcoup-xk}
\end{eqnarray}
Here, the coupling strength of the $j$-th molecule at position $x_j$ to the cavity mode with in-plane momentum $k_x$ is given through
\begin{align}
    g_j(k_x) = -\mu_{01}\cos(\theta_j)\sqrt{\frac{\omega_{\mathrm{cav}}(k_x)}{2\epsilon_0V_{\mathrm{cav}}}},
\end{align}
with molecular electronic transition dipole moment (TDM) at FC point $\mu_{01}$, cavity volume $V_{\mathrm{cav}}$, and angle between the TDM vector and cavity polarization direction $\theta_j$. Cavity and molecular parameters (Tab.~\ref{tab:model-params}) are chosen to closely resemble the Rhodamine microcavity model of Ref.~\cite{sokolovskii_multi-scale_2023}, albeit with a simpler vibrational structure.

Note that we consider a model of an ideal organic crystal, i.e., vibrational frequency, vibronic coupling strength and electronic excitation energy are identical for all molecules, and all TDMs are aligned with the cavity mode polarization direction. The effect of static disorder will be discussed in a future publication along with the details of our quantum-dynamical approach including the ML-MCTDH tree structure for the cavity-molecule wavefunction, and the treatment of non-local couplings between delocalized cavity modes and localized molecular excitations in Eq.~\eqref{eq:hcoup-xk}. In short, transforming the cavity mode basis from momentum to real-space is beneficial as it results in localized couplings and dramatically reduces the number of Hamiltonian terms. 

Within the combined electronic-photonic single-excitation subspace, the Hamiltonian $\hat H$ can be diagonalized analytically to give the dispersion relation of upper and lower polaritonic branches (UPB and LPB) $\omega(k_x)$, as well as the Hopfield coefficients $|\alpha_{\mathrm{LP/UP}}(k_x)|^2$ and $|\beta_{\mathrm{LP/UP}}(k_x)|^2$, i.e. the contributions of cavity modes and molecular excitons to the lower polaritonic and upper polaritonic (LP and UP) states, respectively. Due to their hybrid light-matter nature, EPs can propagate along the in-plane direction of the cavity with group velocity $v_{\mathrm{gr}}^{\mathrm{LP/UP}}=\partial \omega_{\mathrm{LP/UP}}/\partial k_x$. Polariton dispersion, Hopfield coefficients and group velocity play a vital role in the analysis of our quantum dynamical simulations and are shown in Fig.~\ref{fig:Figure_1}b-d. 

\begin{table}
    \centering
    \begin{tabular}{cc}\toprule
        parameter & value \\ \midrule
        $\omega_0$ & 3.81\,eV \\
        $n$   & 1.0 \\
        $\hbar\Omega_R$  & 328\,meV\\
        $\tau_{\mathrm{cav}}$ & 24\,fs\\
        $\omega_m$ & 4.18\,eV\\
        $\omega_{\mathrm{vib}}$ & 74\,meV \\
        $\kappa$ &  0\,meV,  50\,meV,  80\,meV\\ 
        $\Delta x$ & 250\,nm \\
        $F_t$ &  18\,fs \\ 
        $F_x$ &  625\,nm \\
        $t_0$ & 40\,fs\\
        $x_0$ & 2.5\,$\mu$m \\        
        \bottomrule
    \end{tabular}
    \caption{Cavity ($\omega_0$, $n$, $\tau_{\mathrm{cav}}$, $\hbar\Omega_R$), molecular ($\omega_m$,$\omega_{\mathrm{vib}}$,  $\omega_{\mathrm{vib}}$) and laser parameters ($F_t$, $F_x$, $t_0$, $x_0$) relevant to simulations. $F_t$ and $F_x$ denote the temporal and spatial full width at half maximum (FWHM) of the Gaussian laser pulse, respectively. The maxima of temporal and spatial Gaussian envelopes are found at $t_0$ and $x_0$, respectively. }
    \label{tab:model-params}
\end{table}

\section{Results and Discussion}

Experimentally, EP transport is initiated either via off-resonant \cite{balasubrahmaniyam_enhanced_2023,rozenman2018long,myers2018polariton} or resonant excitation \cite{xu_ultrafast_2023,pandya2022tuning,freixanet_-plane_2000,lerario2017high}. In the first case, a focused laser pulse pumps the transition to a high-lying electronic excited state from where the system relaxes incoherently to the S$_1$ state. In our simulations, off-resonant excitation is modeled by instantly promoting a single molecule to the S$_1$ state while all other constituents of the system remain in their respective ground state \cite{sokolovskii_multi-scale_2023}.
In the second case, incidence angle and frequency of the laser pulse target a specific point on the polaritonic dispersion curve, coherently exciting a polaritonic wavepacket centered around a well-defined in-plane momentum $k_x^{(0)}$ and energy $\hbar\omega_L$. We treat such resonant excitation by explicitly incorporating an external laser pulse through a semiclassical dipolar laser-molecule interaction Hamiltonian,  $\hat H_{\mathrm{las}} = -\sum_j \mathcal{E}(x_j,t) \left(\mathrm{e}^{ik_x^{(0)} x_j}\hat\sigma_j^{\dagger}+\mathrm{e}^{-ik_x^{(0)} x_j}\hat\sigma_j\right)$. Temporal and spatial Gaussian envelope functions, $A(t)$ and $B(x_j)$, for the laser field $\mathcal{E}(x_j,t)=A(t)B(x_j)\cos(\omega_L t)$ result in a finite width in energy and momentum space. Laser intensities are chosen low enough to stay within the linear absorption regime, i.e., only one photon is absorbed (cf.~Fig.~S2). All relevant model and laser parameters are summarized in Tab.~\ref{tab:model-params}.

After excitation, we trace the spatiotemporal polariton dynamics by computing the real-time and real-space resolved polaritonic density $|\psi_{\mathrm{pol}}(x_j,t)|^2$ and the individual contributions from the photonic ($|\psi_{\mathrm{phot}}(x_j,t)|^2$) and molecular ($|\psi_{\mathrm{mol}}(x_j,t)|^2$) subsystems, as well as the momentum-resolved photonic density $|\psi_{\mathrm{phot}}(k_{x,p},t)|^2$. These quantities can be readily leveraged from the propagated ML-MCTDH wavefunction by projecting onto localized photonic and molecular excitations, or onto photonic momentum-eigenstates \cite{sokolovskii_multi-scale_2023}.

\subsection{\label{sec:offres}Off-resonant Excitation}
\begin{figure*}
    \centering
    \includegraphics[width=\textwidth]{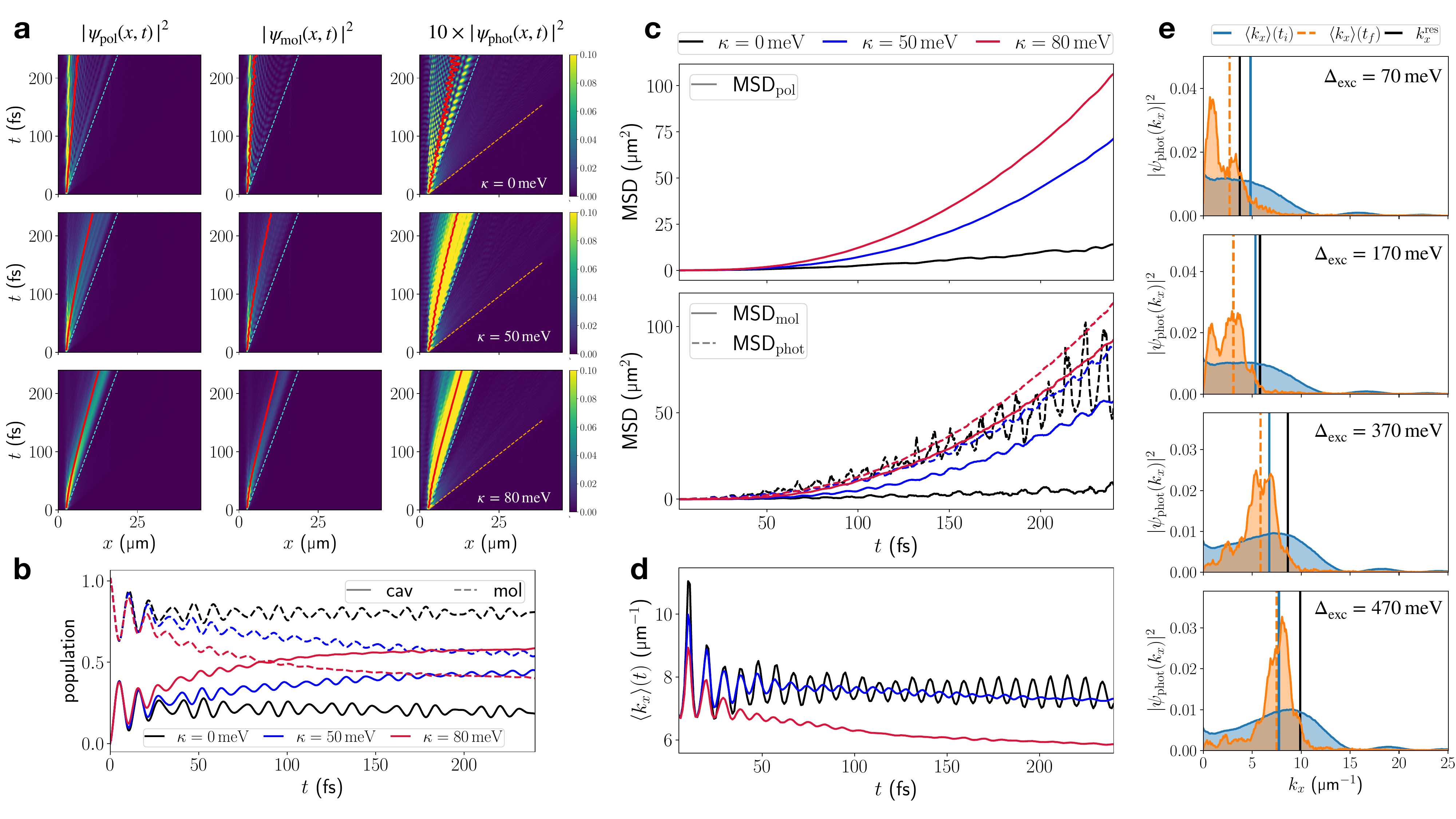}
    \caption{Impact of vibronic coupling on EP transport after off-resonant excitation. (a) Real-space resolved total polaritonic density $|\psi_\mathrm{pol}(x,t)|^2$, molecular and photonic contributions $|\psi_\mathrm{mol}(x,t)|^2$ and $|\psi_\mathrm{phot}(x,t)|^2$, respectively, for vibronic coupling strengths $\kappa=0,50,80$\,meV. Cyan and orange dashed lines indicate linear motion at the maximum group velocities of LPB~(67.9~µm\,ps$^{-1}$)  and UPB~(235~µm\,ps$^{-1}$), respectively. Red lines trace the expectation value of the position operator. (b) Time-dependent population of excited cavity modes (``cav'') and molecular excitons (``mol''). (c) MSDs extracted from real-space resolved densities in (a). (d) Expectation value of cavity mode in-plane momentum for propagating EPs in (a). (e) Initial and final momentum distribution for $\kappa=80$\,meV and various detuning $\Delta_{\mathrm{exc}}$ of molecular electronic excitation energy with respect to cavity photon energy $\omega_0$. The corresponding resonant wavevector $k_x^{\mathrm{res}}$ is indicated by a black vertical line. All distributions in (e) have been normalized.}
    \label{fig:Figure_2}
\end{figure*}

Sudden localized excitation of a single molecule in real-space creates a delocalized excitation in energy-momentum space, affording a superposition of a broad range of LPB and UPB polaritons. Consequently, components with a broad range of group velocities propagate away from the initial off-resonant excitation spot at $x_0=2.5$~µm. 

    Without vibronic coupling ($\kappa=0$), (Fig.~\ref{fig:Figure_2}a, upper panel), a wavefront moving linearly at the maximum LP group velocity $v_{\mathrm{gr}}^{\mathrm{LP},\mathrm{max}}=67.9$~µm\,ps$^{-1}$ is visible in the polaritonic as well as photonic and molecular densities, while zooming in on the photonic real-space density reveals the presence of an additional -- but faint -- wavefront at the maximum UP group velocity which is close to the speed of light ($v_{\mathrm{gr}}^{\mathrm{UP} ,\mathrm{max}}=235$~µm\,ps$^{-1}$).

As such, when vibrations do not couple to the electronic transition, transport after off-resonant excitation is driven mainly by the photonic part of the wavefunction: in Fig.~\ref{fig:Figure_2}a,c mean position and mean square displacement (MSD) of the photonic subsystem exceed those of the molecular subsystem significantly, and a large part of the molecular density remains at the initial excitation spot.

This can be explained by the fact that the initial single-molecule excited state has largest overlap with polaritonic states with high molecular contribution, which are located on the far right of the LPB dispersion, and the far left of the UPB dispersion curve (cf.~Fig.~\ref{fig:Figure_1}c). Thus, the initial molecular excitation mostly spans polaritonic states with vanishing group velocity, as can be seen from Fig.~\ref{fig:Figure_1}d. Fast moving components close to the speed of light originate from high in-plane momentum UPB states which have negligible excitonic contribution (Fig.~\ref{fig:Figure_1}c,d). Therefore,
their signatures are weak and only visible in the photonic density.

The presence of dynamic disorder due to non-zero vibronic coupling $\kappa$ drastically changes the polariton transport dynamics: increasing $\kappa$ in Fig.~\ref{fig:Figure_2}a shifts formerly stationary or slowly propagating components toward the wavefront, which moves at $v_{\mathrm{gr}}^{\mathrm{LP},\mathrm{max}}$. This enhances the mobility substantially, as indicated by the mean polaritonic position and MSD in Fig.~\ref{fig:Figure_2}a,c. Furthermore, the mobility-enhancement occurs dynamically, i.e., the mean group velocity increases over time, which can be seen from the increasing slope of time-dependent mean position $\langle x_{\mathrm{pol}}\rangle(t)$ in Fig.~\ref{fig:Figure_1}a.
The enhancement is more pronounced in the material subsystem, pointing at a vibronically-mediated mechanism that dynamically increases the mobility of
exciton-like polaritons.

This mechanism demands closer analysis. Inspecting the total population of cavity modes and molecular excitonic states in Fig.~\ref{fig:Figure_2}b reveals an increasing photonic content over time for non-zero $\kappa$. For the largest vibronic coupling strength ($\kappa=80$~meV),
the light-matter composition changes from purely excitonic at the beginning of the simulation to a roughly $60/40$
photonic/excitonic composition after approximately 100\,fs.
Moreover, the mean in-plane momentum of cavity modes in Fig.~\ref{fig:Figure_2}d stagnates around 8~µm$^{-1}$ for $\kappa=0$~meV but decreases noticeably to approximately 6~µm$^{-1}$ for $\kappa=80$~meV. These findings suggest vibration-driven relaxation on the LPB as the underlying mechanism for the mobility-enhancement, whereby
population from highly excitonic, near-stationary states is transferred toward more photonic states with large group velocity (cf.~Fig. 1d).

For $\kappa\neq0$, the initial wavepacket formed by off-resonant excitation, no
longer corresponds to a superposition of excitonic-photonic eigenstates.
Instead, intramolecular vibrations couple states on UPB and LPB, inducing
vibration-assisted scattering (VAS) and radiative pumping (RP) processes. VAS and RP have been amply discussed for
organic microcavity EPs
\cite{agranovich_cavity_2003,litinskaya_fast_2004,virgili_ultrafast_2011,coles2011vibrationally,coles2013imaging,tichauer_identifying_2022},
and are relevant for polariton condensation
\cite{byrnes_excitonpolariton_2014,keeling_boseeinstein_2020}.
Highly
excitonic polaritonic states with large in-plane momentum on the LPB can scatter
into a lower-momentum region with increased photonic character and increased
group velocity, resulting in a dynamically increasing mean velocity as indicated by the change in the slope of the red line in Fig.~\ref{fig:Figure_2}a. Scattering
processes on the UPB, on the other hand, do marginally contribute to this
transport-enhancement since fast-propagating, increasingly photonic states are
located at high in-plane momenta and possess high energy, making them hardly
accessible from the off-resonantly excited initial state. Apart from that,
scattering from the UPB to the LPB can contribute, as we discuss in the
next section.

For better understanding of the underlying relaxation process, we inspect the momentum-resolved photonic populations $|\psi_{\mathrm{phot}}(k_x)|^2$ at $t_i=5$\,fs and $t_f=240$\,fs for various detunings $\Delta_{\mathrm{exc}}$ between the excitonic resonance and the fundamental cavity frequency $\omega_0=3.81$\,eV. This tunes the in-plane momentum $k_x^{\mathrm{res}}$ at which cavity mode and exciton are resonant, from low to high in-plane momentum, cf.~Fig.~\ref{fig:Figure_2}e. We find in all cases a clear decrease in population of momentum-states to the right of the resonant in-plane momentum, and an increase to the left. Accordingly, mainly states with $k_x\leq k_x^{\mathrm{res}}$, occasionally referred to as ``bottleneck'' polaritons \cite{myers2018polariton,coles2013imaging}, participate in the vibration-driven relaxation. Moreover, these states do not relax down to $k_x=0$ but rather accumulate in the intermediate region  $0<k_x< k_x^{\mathrm{res}}$ of the LPB where molecular contributions are still appreciable and group velocities are high (cf.~Fig.~\ref{fig:Figure_1}c,d).

Further relaxation is hindered by the decreasing efficiency of exciton-mediated scattering when the photonic content of involved states becomes large \cite{litinskaya_fast_2004, coles2013imaging, laitz2023uncovering}. This effect is known as the ``phonon bottleneck'' in the context of EP condensation \cite{byrnes_excitonpolariton_2014,sanvitto_road_2016}. Note that the wavevector-resolved photonic population cannot capture highly excitonic polaritons due to their vanishing photonic contribution. Nonetheless, the disappearing stationary component at $x_0$ of $|\psi_{\mathrm{mol}}(x,t)|^2$ for $\kappa=50$\,meV and $\kappa=80$\,meV in Fig.~\ref{fig:Figure_2}a indicates that they undergo relaxation on the LPB as well.

\subsection{\label{sec:onres}Resonant Excitation}

\begin{figure*}
    \centering
    \includegraphics[width=0.95\textwidth]{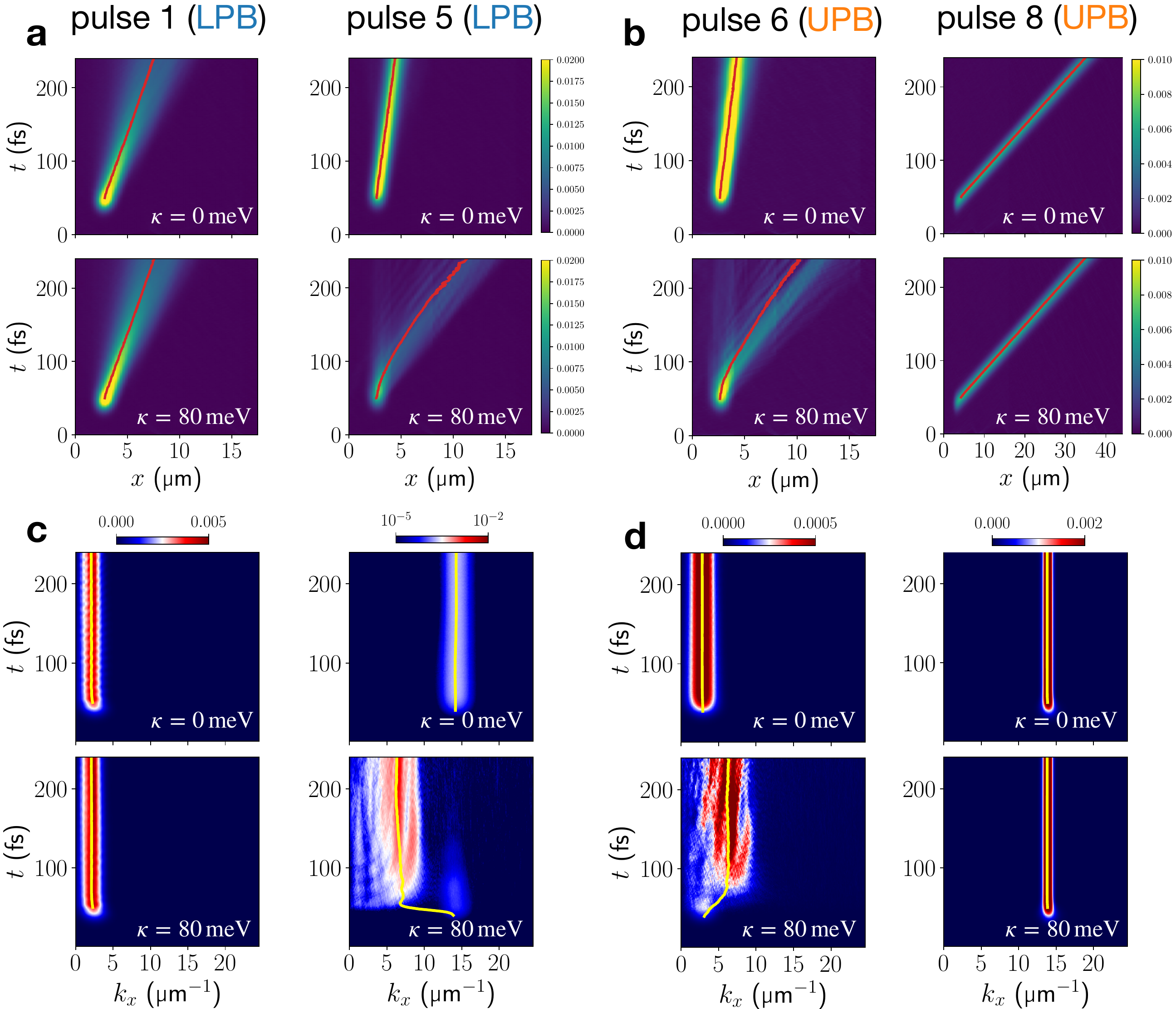}
    \caption{Transport enhancement through vibronically-mediated relaxation of resonantly excited EP wavepackets. (a) Real-space resolved total polaritonic density $|\psi_\mathrm{pol}(x,t)|^2$ for resonant-excitation to the LPB by pulse 1 and 5, at vibronic coupling strengths $\kappa=0$\,meV and $\kappa=80$\,meV. (b) Same as (a) but targeting the UPB by pulses 6 and 8. Red lines indicate the expectation value of the position operator. The momentum-resolved cavity mode populations corresponding to (a) and (b) are shown in (c) and (d), respectively. Yellow lines indicate the expectation value of the cavity-mode momentum operator. }
    \label{fig:Figure_3}
\end{figure*}

In the following, we investigate the relevance of molecular vibronic coupling in
a wide range of resonant excitation scenarios targeting LPB and UPB. To this
end, eight laser pulses are considered with pulse parameters (except photon
energy and incidence angle, which are varied) given in
Tab.~\ref{tab:model-params}. The resonantly targeted points in the
$[\omega(k_x),k_x]$ plane are shown in
Fig.~\ref{fig:Figure_1}b.
First, we discuss two limiting cases on
each branch: a highly photonic and a highly excitonic wavepacket,
which, on the LPB, are found at small and large wavevectors, respectively.  The
situation inverts on the UPB, where the photonic content increases towards large
wavevectors.

The LPB dispersion relation at small wavevectors is parabolic such that
coherently-excited wavepackets in this region display ballistic transport and
dispersive behavior \cite{agranovich_nature_2007}. Our full-quantum dynamical
results reproduce this for resonant excitation targeted at
$k_x^{(0)}=2.75$\,µm$^{-1}$ (pulse 1) on the LPB, corresponding to an exciton
content of 13.8\% (cf.~Fig.~\ref{fig:Figure_1}c). In Fig.~\ref{fig:Figure_3}a
the mean position of the polaritonic density moves linearly, corresponding to
quadratic growth of the MSD, while the width of the wavepacket spreads visibly
during propagation in a lossless cavity. Polaritonic densities in both real and
momentum space are identical for $\kappa=0$\,meV and $\kappa=80$\,meV
(Fig.~\ref{fig:Figure_3}a,c); the wavevector-resolved cavity-mode population
stays constant after the pulse, showing no signatures of relaxation
(Fig.~\ref{fig:Figure_3}c). As discussed for the off-resonant case, the
efficiency of vibration-mediated scattering processes decreases significantly
towards the small-wavevector region on the LPB due to increasing photonic
content. Consequently, the ballistic dispersive transport is unaffected by
vibronic coupling in this regime.

Strong differences are found when the laser pulse addresses the highly excitonic
region on the LPB, exciting at $k_x^{(0)}=14.5$\,µm$^{-1}$ (pulse 5)
corresponding to an exciton content of $\sim$90\%. For $\kappa=0$\,meV transport
proceeds ballistically, Fig.~\ref{fig:Figure_3}a, at low group velocity, but
dispersion during propagation is less pronounced due to the non-parabolic, flat
LPB dispersion relation in the large in-plane momentum region. Vibronic coupling
significantly changes this behavior. Again, strong vibronic
transport-enhancement becomes apparent from the polaritonic real-space density
and mean position in Fig.~\ref{fig:Figure_3}a. Wavevector-resolved cavity
populations in Fig.~\ref{fig:Figure_3}c reveals rapid relaxation of high
in-plane momentum components of the LPB wavepacket to faster-propagating
polaritons close to the ``bottleneck region'' below
$k_x^{\mathrm{res}}=8.64$~µm$^{-1}$ as the origin of the enhancement. These
findings verify the participation of highly-excitonic LPB states in the
vibration-mediated relaxation as discussed for off-resonantly excited
polaritons.

Targeting the corresponding UPB states with pulses~6 and 8 results in an inverted situation: Now, excitation at large wavevectors
affords a highly photonic EP wavepacket with large group velocity which is
shielded from molecular vibronic effects through its low exciton content of
$\sim$7\%.  Consequently, we find ballistic transport independent of the
vibronic coupling strength in Fig.~\ref{fig:Figure_3}b. In contrast,
low-wavevector (pulse~6) excitation creates a predominantly excitonic EP
wavepacket with low group velocity. Again, vibronic coupling results in
relaxation of the wavepacket on the UPB to the ``bottleneck region'' in
Fig.~\ref{fig:Figure_3}d which follows a similar fate as the LPB wavepacket
excited by pulse 5 in Fig.~\ref{fig:Figure_3}c. As a result, the propagation
velocity is significantly enhanced, producing very similar trajectories for
initially highly excitonic wavepackets on the LPB and UPB (pulse 5 in
Fig.~\ref{fig:Figure_3}a and pulse 8 in Fig.~\ref{fig:Figure_3}b). This suggests
that vibronic interactions within molecules can efficiently funnel population
towards the high-mobility region on the LPB close to $k_x^{\mathrm{res}}$ from
both UP and LP states.

\begin{figure*}
    \centering
    \includegraphics[width=\textwidth]{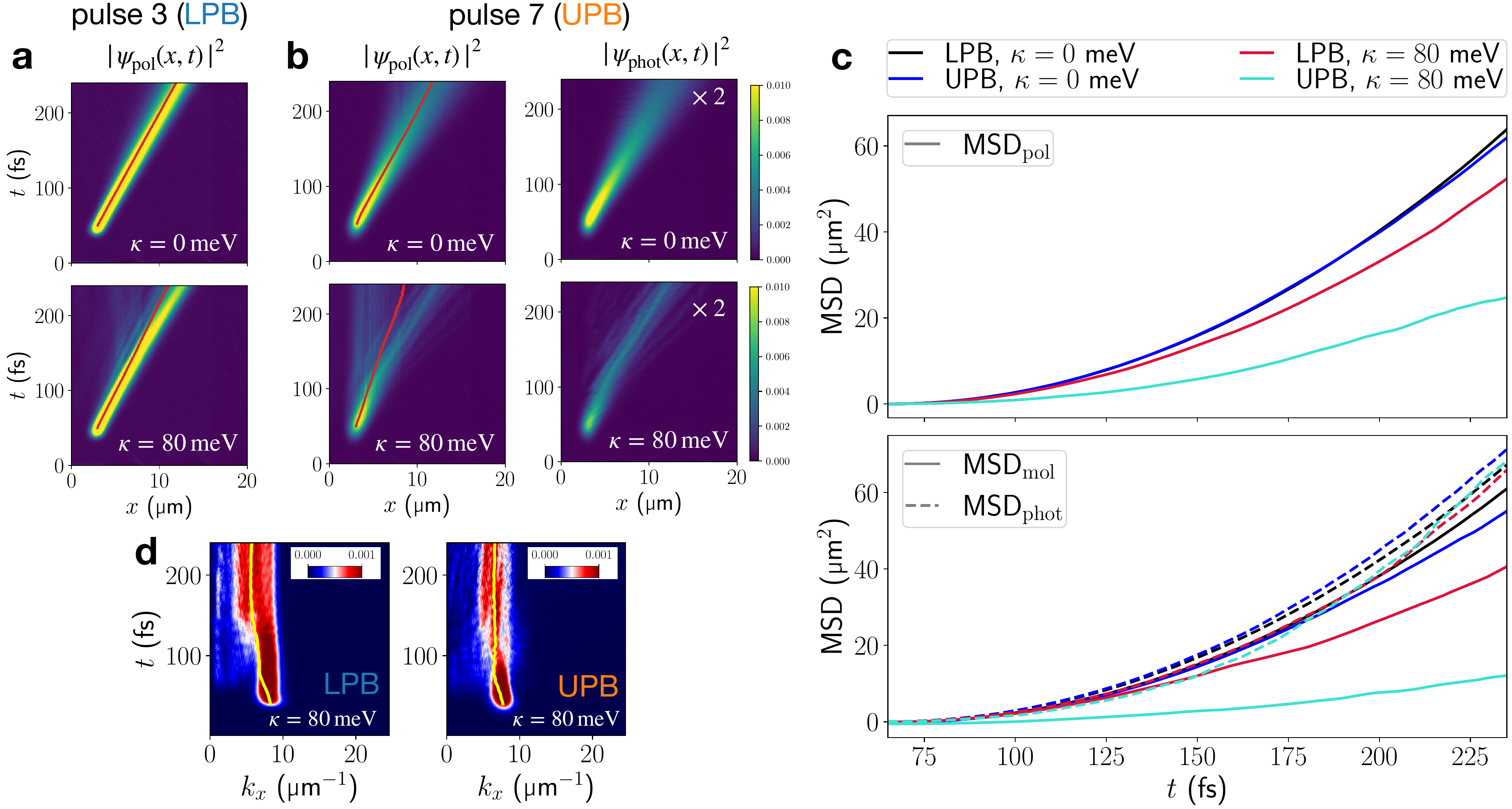}
    \caption{EP propagation ensuing excitation at resonant wavevector. (a) Real-space resolved total polaritonic density $|\psi_\mathrm{pol}(x,t)|^2$ for resonant-wavevector excitation to the LPB. (b) Real-space resolved total polaritonic density $|\psi_\mathrm{pol}(x,t)|^2$ and photonic contribution $|\psi_\mathrm{phot}(x,t)|^2$ for resonant-wavevector excitation to the UPB. Red lines indicate the expectation value of the position operator. (c) MSDs extracted from polaritonic real-space resolved densities in (a) and (b). (d) Momentum-resolved cavity mode populations corresponding to EP propagations at $\kappa=80$\,meV in (a) and (b). Yellow lines indicate the expectation value of the cavity-mode momentum operator.}
    \label{fig:Figure_4}
\end{figure*}

 Next, we target the high-mobility region directly by excitation to the UPB and
 LPB at the resonant wavevector $k_x^{\mathrm{res}}$ (pulses~3 and 7 in
 Fig.~\ref{fig:Figure_1}b). Although possessing the same nominal group velocity
 and equal light-matter composition in the absence of vibronic coupling, UP and
 LP states at $k_x^{\mathrm{res}}$ undergo distinct relaxation dynamics, both
 resulting in a slow-down of polariton propagation when vibronic coupling is
 included, as seen in Fig.~\ref{fig:Figure_4}c. This is in stark contrast to the
 previously discussed highly excitonic UP and LP states which both follow
 similar relaxation pathways and exhibit vibronically-enhanced transport.
  After UP excitation, the slow-down effect results in the onset of diffusive transport, which is indicated by the linear growth of the MSD in Fig.~\ref{fig:Figure_4}c. As seen in Fig.~\ref{fig:Figure_4}b, vibronic coupling leads to the appearance of non-propagating ``vertical''
 features which lack any contribution from the
 cavity modes. 
 
 Therefore, we can attribute the slow-down effect for UP
 excitation to non-radiative decay from the UPB towards the dark states driven
 by vibronic motion. Polariton dynamics around the resonant wavevector can be
 regarded in close analogy to that of many molecules electronically coupled to a
 single resonant cavity mode. In that case, vibronic interactions in molecules
 have been found to support fast and efficient decay to the dark-state manifold
 from the UP state \cite{vendrell_collective_2018,feist_polaritonic_2018}.

 For LP excitation at $k_x^{\mathrm{res}}$, such vertical features are not
 visible in the polaritonic density (cf. Fig.~\ref{fig:Figure_4}a). Instead,
 fractions with lower group-velocities depart from propagating wavefront over
 time, indicating population transfer to polaritonic states with lower group
 velocities. Accordingly, intraband relaxation on the LPB dominates the dynamics
 after LP excitation which can be traced through the wavevector-resolved
 photonic density in Fig.~\ref{fig:Figure_4}d. While this mechanism generally
 enhances the transport as observed in this study for off-resonant and resonant
 excitation schemes, it can only access slower-propagating states when starting
 at the resonant wavevector, i.e. close to the maximum of the LPB group velocity
 (cf.~Fig.~\ref{fig:Figure_1}d).

Summarizing, our fully quantum-dynamical simulations of EP dynamics in a
lossless cavity show that vibronic motion can strongly impact polariton
transport depending on the initially targeted state. Vibronic interactions
provide relaxation channels along and among polaritonic branches and dark-states
for appreciable exciton content ($\geq 20\%$) of the targeted EP state.
Generally, fast-propagating states close to the polariton bottleneck can be
accessed efficiently from exciton-like UPB and LPB states. This results in a
vibronic enhancement of EP transport. However, close to the inflection point of
LPB and UPB trapping within the dark-state manifold and low-group velocity EP
states close to $k_x=0$ become possible, resulting in a slow-down of EP
transport.

\subsection{\label{sec:decay}Impact of radiative decay}
Finally, we introduce cavity losses to the cavity-molecule system since current
Fabry-Pérot cavity setups based on high-reflectivity metallic mirrors have
radiative lifetimes of typically 10\,fs to 30\,fs due to cavity photon leakage
to the electromagnetic continuum. This fundamentally limits the lifetime of
polaritonic state, and thus has substantial impact on the transport properties
of cavity-molecule systems. We include spontaneous cavity losses in ML-MCTDH
propagations through non-Hermitian damping terms
$-\sum_{k_x}\mathrm{i}\frac{\Gamma_{k_x}}{2}\hat{a}_{k_x}^{\dagger}\hat{a}_{k_x}$
\cite{ulusoy_dynamics_2020,tichauer_tuning_2023}. A uniform decay constant
$\Gamma_C=1/\tau_C$ is assumed for all cavity modes, with apparent cavity
lifetime $\tau_C=24$\,fs. As pointed out in earlier works, this ansatz is
equivalent to propagating the density matrix according to a Lindblad master
equation if the system stays within the electronic-photonic single-excitation subspace \cite{felicetti_photoprotecting_2020,fabri_coupling_2024}. In our simulations, coupling strengths are sufficiently low and external pulses sufficiently short and weak such that this condition is met (Fig.~S2). When the norm of the polaritonic wavepacket has decreased below 1\% of the initial laser-excited population, the wavepacket is considered to have decayed and MSDs are not computed beyond this point.

\begin{figure*}
    \centering
    \includegraphics[width=\textwidth]{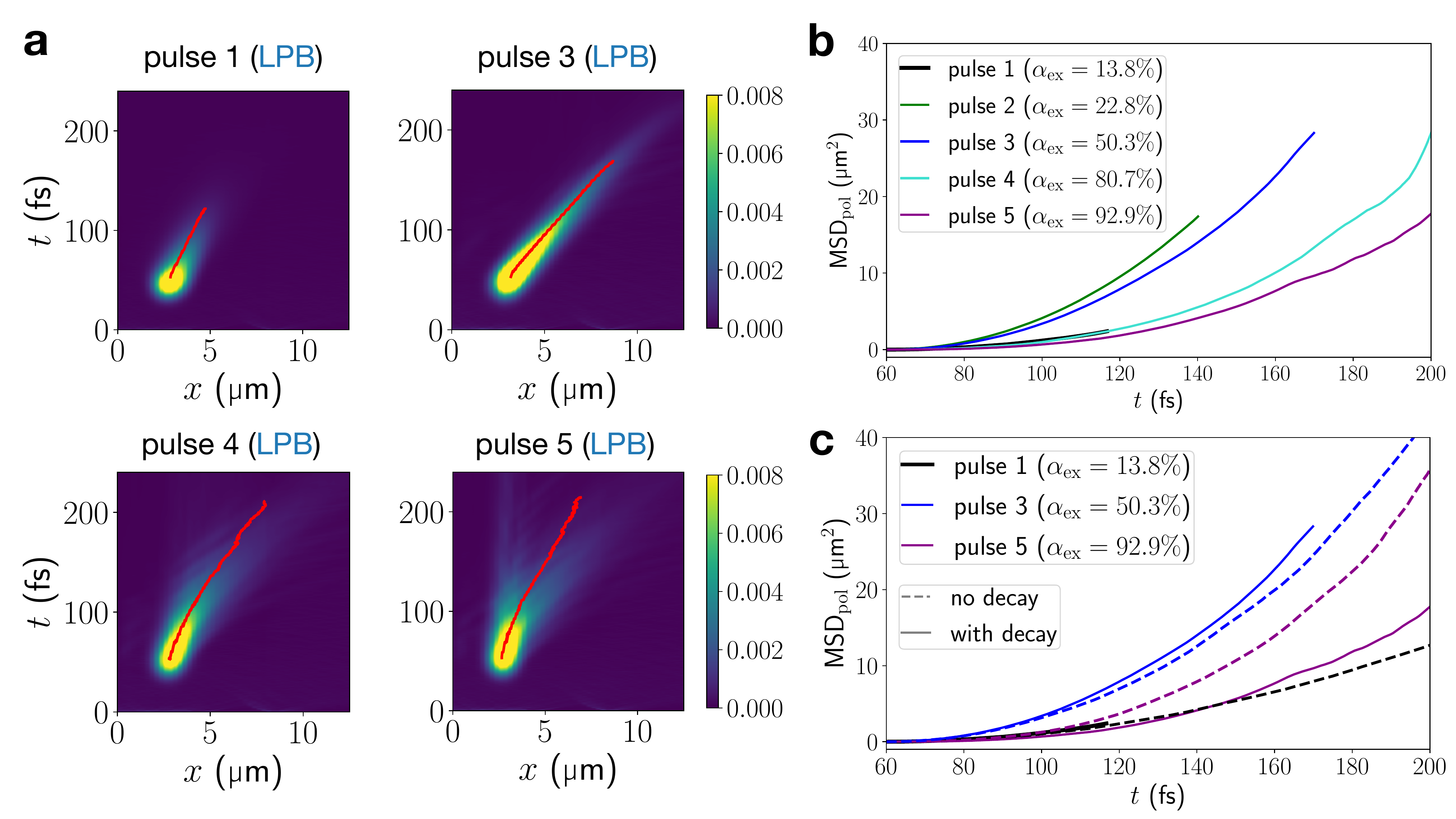}
    \caption{Impact of radiative decay on polariton transport. (a) Real-space resolved polaritonic densities  $|\psi_\mathrm{pol}(x,t)|^2$ for $\kappa=80$\,meV and three resonantly excited LPB wavepackets in a lossy cavity with $\tau_C=24$\,fs. (b) MSD of propagating EP for a vibronic coupling strength of $\kappa=80$\,meV. EPs have been excited resonantly to five different points at the LPB by pulses 1-5; exciton contents $\alpha_{\mathrm{ex}}$ at these targeted points are given in the legend. Red lines indicate the expectation value of the position operator. (c) Comparison of MSDs for pulses 1,3,5 in the presence and absence of radiative decay.}
    \label{fig:Figure_5}
\end{figure*}

We compare the spatiotemporal evolution of five coherently excited LP
wavepackets with increasing exciton content, tuning from the photonic (pulse 1)
to the excitonic regime (pulse 5). While high photonic character of the
initially excited polaritonic wavepacket protects transport against the impact
of vibronic coupling, it makes EP transport highly vulnerable towards radiative
decay in low-Q cavities. Since the emission rate is proportional to the photon
number, the polaritonic density rapidly decays in a lossy cavity.
This limits the
observable transport to less than 100\,fs after the pulse maximum, as
seen in
Fig.~\ref{fig:Figure_5}a,b.

Moreover, our simulations in a lossy cavity indicate a significantly prolonged
propagation when targeting the high in-plane momentum region of the LPB in
Fig.~\ref{fig:Figure_5}b. Due to their excitonic nature, the impact of cavity
losses on their lifetime is mitigated compared to the low-wavevector regime.
Yet, vibronic interactions connect them to the more photonic bottleneck region
of the LPB. Thus, exciton-like large-wavevector LP states can act as a reservoir
from which population is fed to the fast-propagating polaritonic manifold. This
leads to long-lived and efficient transport due to -- not despite of -- vibronic
coupling. Our results are in line with a recent experiment
\cite{xu_ultrafast_2023} which could support the exciton reservoir hypothesis
\cite{virgili_ultrafast_2011, bhuyan2024effect} as well.

In a lossy cavity, the interplay of vibronic effects with the varying group velocities and lifetimes along the EP dispersion curve creates distinct transport regimes on the LPB. 
The low-wavevector region (pulse~1) is characterized by low to moderate group velocities and rapid radiative decay, resulting in short-lived and short-distance polariton transport. Entering the bottleneck region (pulse~2), the group velocity has a maximum but propagation duration is still strongly limited due to appreciable photonic content. Around the resonant wavevector (pulse~3), group velocities are still high and the increased excitonic content protects polaritons from cavity losses, enabling long-distance transport. Our simulations indicate that vibronic coupling has only minor impact on polariton motion in the regimes left of the resonant wavevector.

Remarkably, transport properties of lower-band polaritons to the right of the resonant wavevector (pulse~4) are improved in the presence of vibronic coupling by combining their long lifetimes and efficient access to the high-mobility bottleneck region of the LPB. This is illustrated by comparing polariton transport ensuing pulse 1 and pulse 4 (Fig.~\ref{fig:Figure_5}). Although both pulses target LPB states with very similar group velocities, the initially exciton-like wavepacket (pulse 4) propagates much longer and further than the quickly decaying photon-like wavepacket (pulse~1). When targeting even higher wavevectors, the vanishing photonic contribution and small group velocity begin to dominate (pulse~5). Overall, this affords an optimal transport regime on the LPB between exciton contents of approximately 50\% and 80\%. Initial photoexcitation to this region harnesses beneficial vibronic effects and favorable transport properties set by the polariton dispersion relation.
As seen in Fig.~\ref{fig:Figure_5}b, LPB wavepackets with $\sim$50\% (pulse 3) and  $\sim$80\% (pulse 4) exciton content display the longest propagation distances after 200\,fs.

\section{Discussion and Outlook}

We have simulated exciton-polariton transport in a coupled multimode cavity, treating molecular vibrations, electronic states
and cavity photons as quantum degrees of freedom. To our knowledge,
fully-quantum dynamical simulations of EP transport are unprecedented. In this
work, they are made possible by state-of-the art quantum-dynamical methods based
on a compact tree-tensor network wavefunction ansatz.

Our results show a strong dependence of polariton transport dynamics on vibronic
interactions within the molecules: intramolecular vibrations mediate
both inter- and intraband relaxation dynamics whose impact on EP transport
depends strongly on the
photon energy and wavevector of the incoming laser light.
When targeting exciton-like polaritons, vibronic coupling leads to efficient population transfer toward fast-propagating polaritons at the bottleneck of the LPB. In lossy cavities, this prolongs the lifetime of bottleneck-polaritons since the bottleneck-region is populated transiently from long-lived excitonic states. This enables long-range polariton transport on the LPB even in the presence of cavity photon decay.%
Thereby, our simulations highlight the major importance of molecular
photophysical properties for efficient EP transport.

Previously, multi-scale semi-classical dynamics simulations
\cite{sokolovskii_multi-scale_2023} found enhanced mobility at 300\,K when
comparing with simulations at 0\,K (frozen nuclei), both under off-resonant
pump conditions. This highlighted the role of vibrations in transferring
population from stationary bare excitonic states to propagating bright
polaritonic states.
In these semi-classical simulations, ballistic propagation at short times is
always superseded by the onset of diffusive transport after about 100~fs (cf.
Fig.~3 in Ref.~\cite{sokolovskii_multi-scale_2023}).

Our quantum dynamics simulations indicate ballistic transport for all vibronic
coupling strengths when exciting the LPB. Vibronic coupling provides mobility to
the wavepacket in k-space, whereas cavity losses decrease the population of
highly photonic components, thus effectively decreasing the propagation velocity
(cf. Fig.~\ref{fig:Figure_5}c), in agreement with the simulations in
Ref.~\cite{sokolovskii_multi-scale_2023}.
We observed the onset of diffusive transport only when resonantly exciting the
UPB. In this case, the wavepacket quickly decays into the dark states manifold
due to vibronic coupling. This leads to an apparent polariton wavepacket contraction in the presence of cavity losses (Fig.~S3), confirming the analysis based on semi-classical simulations in Ref.~\cite{tichauer_tuning_2023}: the stationary dark-state population persists while the propagating fraction of the wavepacket decays quickly due to its higher photonic content.
Experimental findings on EP transport vary strongly \cite{freixanet_-plane_2000, rozenman2018long,xu_ultrafast_2023,lerario2017high}
suggesting preserved ballistic flow or slower-than-expected diffusive
transport, while a unified explanation is still lacking.
These ambiguities point at a more complex mechanism behind exciton-polariton
propagation which demands a clear distinction between individual
effects. More precisely, dynamic disorder due to vibronic coupling, and due to
thermal motion of the nuclei in the ground state, as well as static disorder due
to inhomogeneities in the solid phase can have distinct individual and
cooperative effects on transport.
Vibronic, thermal, and static disorder effects can be cleanly separated in a
fully-quantum description, but become intermingled in semi-classical approaches.

Our rigorous treatment of vibronic coupling indicates that electron-phonon
scattering due to intramolecular vibronic coupling alone cannot fully account
for velocity reduction and the transition from ballistic to diffusive transport.
Recent experiments have already pointed out the importance of thermal effects
in EP transport. At very low temperature (5~K) the transport remains ballistic,
as in our simulations at 0~K. The onset of diffusive transport is observed at
room temperature \cite{xu_ultrafast_2023}.%
Thus, we speculate that vibronic effects are responsible for enabling the
mobility of the EP wavepacket in k-space, where it accumulates in the bottleneck
region, whereas thermal disorder plays a critical role in
reducing propagation velocity and limiting the ballistic flow
observed in experiments.

Nonetheless, the number of simulated molecules within the quantization length
is smaller than in experiments \cite{bhuyan2024effect} 
and in previous semi-classical calculations \cite{sokolovskii_multi-scale_2023}.
This might underestimate the impact of dark states in our simulations.
Participation of the dark-state manifold may be thermally activated as suggested
in Refs.~\cite{balasubrahmaniyam_enhanced_2023,sandik2024cavity}, and can result
in the onset of diffusive transport \cite{sokolovskii_multi-scale_2023}. The
exact mechanism for the emergence of diffusive transport at moderate to high
temperatures is still a matter of debate.

The insights from this work pave the way toward a detailed mechanistic picture
of polariton transport and relaxation in increasingly complex systems from
first-principles.
Future work in our laboratory will focus on the role of static disorder, finite
temperature, and the impact of dark states. Developing a microscopic
understanding of polariton propagation is crucial for identifying materials with
properties suited for efficient and functional polaritonic devices.


\begin{thebibliography}{50}%
\makeatletter
\providecommand \@ifxundefined [1]{%
 \@ifx{#1\undefined}
}%
\providecommand \@ifnum [1]{%
 \ifnum #1\expandafter \@firstoftwo
 \else \expandafter \@secondoftwo
 \fi
}%
\providecommand \@ifx [1]{%
 \ifx #1\expandafter \@firstoftwo
 \else \expandafter \@secondoftwo
 \fi
}%
\providecommand \natexlab [1]{#1}%
\providecommand \enquote  [1]{``#1''}%
\providecommand \bibnamefont  [1]{#1}%
\providecommand \bibfnamefont [1]{#1}%
\providecommand \citenamefont [1]{#1}%
\providecommand \href@noop [0]{\@secondoftwo}%
\providecommand \href [0]{\begingroup \@sanitize@url \@href}%
\providecommand \@href[1]{\@@startlink{#1}\@@href}%
\providecommand \@@href[1]{\endgroup#1\@@endlink}%
\providecommand \@sanitize@url [0]{\catcode `\\12\catcode `\$12\catcode
  `\&12\catcode `\#12\catcode `\^12\catcode `\_12\catcode `\%12\relax}%
\providecommand \@@startlink[1]{}%
\providecommand \@@endlink[0]{}%
\providecommand \url  [0]{\begingroup\@sanitize@url \@url }%
\providecommand \@url [1]{\endgroup\@href {#1}{\urlprefix }}%
\providecommand \urlprefix  [0]{URL }%
\providecommand \Eprint [0]{\href }%
\providecommand \doibase [0]{https://doi.org/}%
\providecommand \selectlanguage [0]{\@gobble}%
\providecommand \bibinfo  [0]{\@secondoftwo}%
\providecommand \bibfield  [0]{\@secondoftwo}%
\providecommand \translation [1]{[#1]}%
\providecommand \BibitemOpen [0]{}%
\providecommand \bibitemStop [0]{}%
\providecommand \bibitemNoStop [0]{.\EOS\space}%
\providecommand \EOS [0]{\spacefactor3000\relax}%
\providecommand \BibitemShut  [1]{\csname bibitem#1\endcsname}%
\let\auto@bib@innerbib\@empty
\bibitem [{\citenamefont {Weisbuch}\ \emph {et~al.}(1992)\citenamefont
  {Weisbuch}, \citenamefont {Nishioka}, \citenamefont {Ishikawa},\ and\
  \citenamefont {Arakawa}}]{weisbuch_observation_1992}%
  \BibitemOpen
  \bibfield  {author} {\bibinfo {author} {\bibfnamefont {C.}~\bibnamefont
  {Weisbuch}}, \bibinfo {author} {\bibfnamefont {M.}~\bibnamefont {Nishioka}},
  \bibinfo {author} {\bibfnamefont {A.}~\bibnamefont {Ishikawa}},\ and\
  \bibinfo {author} {\bibfnamefont {Y.}~\bibnamefont {Arakawa}},\ }\bibfield
  {title} {\bibinfo {title} {Observation of the coupled exciton-photon mode
  splitting in a semiconductor quantum microcavity},\ }\href
  {https://doi.org/10.1103/PhysRevLett.69.3314} {\bibfield  {journal} {\bibinfo
   {journal} {Phys. Rev. Lett.}\ }\textbf {\bibinfo {volume} {69}},\ \bibinfo
  {pages} {3314} (\bibinfo {year} {1992})}\BibitemShut {NoStop}%
\bibitem [{\citenamefont {Bhuyan}\ \emph {et~al.}(2023)\citenamefont {Bhuyan},
  \citenamefont {Mony}, \citenamefont {Kotov}, \citenamefont {Castellanos},
  \citenamefont {Gómez~Rivas}, \citenamefont {Shegai},\ and\ \citenamefont
  {Börjesson}}]{bhuyan_rise_2023}%
  \BibitemOpen
  \bibfield  {author} {\bibinfo {author} {\bibfnamefont {R.}~\bibnamefont
  {Bhuyan}}, \bibinfo {author} {\bibfnamefont {J.}~\bibnamefont {Mony}},
  \bibinfo {author} {\bibfnamefont {O.}~\bibnamefont {Kotov}}, \bibinfo
  {author} {\bibfnamefont {G.~W.}\ \bibnamefont {Castellanos}}, \bibinfo
  {author} {\bibfnamefont {J.}~\bibnamefont {Gómez~Rivas}}, \bibinfo {author}
  {\bibfnamefont {T.~O.}\ \bibnamefont {Shegai}},\ and\ \bibinfo {author}
  {\bibfnamefont {K.}~\bibnamefont {Börjesson}},\ }\bibfield  {title}
  {\bibinfo {title} {The {Rise} and {Current} {Status} of {Polaritonic}
  {Photochemistry} and {Photophysics}},\ }\href
  {https://doi.org/10.1021/acs.chemrev.2c00895} {\bibfield  {journal} {\bibinfo
   {journal} {Chem. Rev.}\ }\textbf {\bibinfo {volume} {123}},\ \bibinfo
  {pages} {10877} (\bibinfo {year} {2023})}\BibitemShut {NoStop}%
\bibitem [{\citenamefont {Khazanov}\ \emph {et~al.}(2023)\citenamefont
  {Khazanov}, \citenamefont {Gunasekaran}, \citenamefont {George},
  \citenamefont {Lomlu}, \citenamefont {Mukherjee},\ and\ \citenamefont
  {Musser}}]{khazanov_embrace_2023}%
  \BibitemOpen
  \bibfield  {author} {\bibinfo {author} {\bibfnamefont {T.}~\bibnamefont
  {Khazanov}}, \bibinfo {author} {\bibfnamefont {S.}~\bibnamefont
  {Gunasekaran}}, \bibinfo {author} {\bibfnamefont {A.}~\bibnamefont {George}},
  \bibinfo {author} {\bibfnamefont {R.}~\bibnamefont {Lomlu}}, \bibinfo
  {author} {\bibfnamefont {S.}~\bibnamefont {Mukherjee}},\ and\ \bibinfo
  {author} {\bibfnamefont {A.~J.}\ \bibnamefont {Musser}},\ }\bibfield  {title}
  {\bibinfo {title} {Embrace the darkness: {An} experimental perspective on
  organic exciton–polaritons},\ }\href {https://doi.org/10.1063/5.0168948}
  {\bibfield  {journal} {\bibinfo  {journal} {Chemical Physics Reviews}\
  }\textbf {\bibinfo {volume} {4}},\ \bibinfo {pages} {041305} (\bibinfo {year}
  {2023})}\BibitemShut {NoStop}%
\bibitem [{\citenamefont {Jiang}\ \emph {et~al.}(2022)\citenamefont {Jiang},
  \citenamefont {Ren}, \citenamefont {Yan}, \citenamefont {Yao},\ and\
  \citenamefont {Zhao}}]{jiang_excitonpolaritons_2022}%
  \BibitemOpen
  \bibfield  {author} {\bibinfo {author} {\bibfnamefont {Z.}~\bibnamefont
  {Jiang}}, \bibinfo {author} {\bibfnamefont {A.}~\bibnamefont {Ren}}, \bibinfo
  {author} {\bibfnamefont {Y.}~\bibnamefont {Yan}}, \bibinfo {author}
  {\bibfnamefont {J.}~\bibnamefont {Yao}},\ and\ \bibinfo {author}
  {\bibfnamefont {Y.~S.}\ \bibnamefont {Zhao}},\ }\bibfield  {title} {\bibinfo
  {title} {Exciton‐{Polaritons} and {Their} {Bose}–{Einstein} {Condensates}
  in {Organic} {Semiconductor} {Microcavities}},\ }\href
  {https://doi.org/10.1002/adma.202106095} {\bibfield  {journal} {\bibinfo
  {journal} {Advanced Materials}\ }\textbf {\bibinfo {volume} {34}},\ \bibinfo
  {pages} {2106095} (\bibinfo {year} {2022})}\BibitemShut {NoStop}%
\bibitem [{\citenamefont {Byrnes}\ \emph {et~al.}(2014)\citenamefont {Byrnes},
  \citenamefont {Kim},\ and\ \citenamefont
  {Yamamoto}}]{byrnes_excitonpolariton_2014}%
  \BibitemOpen
  \bibfield  {author} {\bibinfo {author} {\bibfnamefont {T.}~\bibnamefont
  {Byrnes}}, \bibinfo {author} {\bibfnamefont {N.~Y.}\ \bibnamefont {Kim}},\
  and\ \bibinfo {author} {\bibfnamefont {Y.}~\bibnamefont {Yamamoto}},\
  }\bibfield  {title} {\bibinfo {title} {Exciton–polariton condensates},\
  }\href {https://doi.org/10.1038/nphys3143} {\bibfield  {journal} {\bibinfo
  {journal} {Nature Phys}\ }\textbf {\bibinfo {volume} {10}},\ \bibinfo {pages}
  {803} (\bibinfo {year} {2014})}\BibitemShut {NoStop}%
\bibitem [{\citenamefont {Su}\ \emph {et~al.}(2020)\citenamefont {Su},
  \citenamefont {Ghosh}, \citenamefont {Wang}, \citenamefont {Liu},
  \citenamefont {Diederichs}, \citenamefont {Liew},\ and\ \citenamefont
  {Xiong}}]{su_observation_2020}%
  \BibitemOpen
  \bibfield  {author} {\bibinfo {author} {\bibfnamefont {R.}~\bibnamefont
  {Su}}, \bibinfo {author} {\bibfnamefont {S.}~\bibnamefont {Ghosh}}, \bibinfo
  {author} {\bibfnamefont {J.}~\bibnamefont {Wang}}, \bibinfo {author}
  {\bibfnamefont {S.}~\bibnamefont {Liu}}, \bibinfo {author} {\bibfnamefont
  {C.}~\bibnamefont {Diederichs}}, \bibinfo {author} {\bibfnamefont {T.~C.~H.}\
  \bibnamefont {Liew}},\ and\ \bibinfo {author} {\bibfnamefont
  {Q.}~\bibnamefont {Xiong}},\ }\bibfield  {title} {\bibinfo {title}
  {Observation of exciton polariton condensation in a perovskite lattice at
  room temperature},\ }\href {https://doi.org/10.1038/s41567-019-0764-5}
  {\bibfield  {journal} {\bibinfo  {journal} {Nat. Phys.}\ }\textbf {\bibinfo
  {volume} {16}},\ \bibinfo {pages} {301} (\bibinfo {year} {2020})}\BibitemShut
  {NoStop}%
\bibitem [{\citenamefont {Daskalakis}\ \emph {et~al.}(2015)\citenamefont
  {Daskalakis}, \citenamefont {Maier},\ and\ \citenamefont
  {Kéna-Cohen}}]{daskalakis_spatial_2015}%
  \BibitemOpen
  \bibfield  {author} {\bibinfo {author} {\bibfnamefont {K.}~\bibnamefont
  {Daskalakis}}, \bibinfo {author} {\bibfnamefont {S.}~\bibnamefont {Maier}},\
  and\ \bibinfo {author} {\bibfnamefont {S.}~\bibnamefont {Kéna-Cohen}},\
  }\bibfield  {title} {\bibinfo {title} {Spatial {Coherence} and {Stability} in
  a {Disordered} {Organic} {Polariton} {Condensate}},\ }\href
  {https://doi.org/10.1103/PhysRevLett.115.035301} {\bibfield  {journal}
  {\bibinfo  {journal} {Phys. Rev. Lett.}\ }\textbf {\bibinfo {volume} {115}},\
  \bibinfo {pages} {035301} (\bibinfo {year} {2015})}\BibitemShut {NoStop}%
\bibitem [{\citenamefont {Plumhof}\ \emph {et~al.}(2014)\citenamefont
  {Plumhof}, \citenamefont {Stöferle}, \citenamefont {Mai}, \citenamefont
  {Scherf},\ and\ \citenamefont {Mahrt}}]{plumhof_room-temperature_2014}%
  \BibitemOpen
  \bibfield  {author} {\bibinfo {author} {\bibfnamefont {J.~D.}\ \bibnamefont
  {Plumhof}}, \bibinfo {author} {\bibfnamefont {T.}~\bibnamefont {Stöferle}},
  \bibinfo {author} {\bibfnamefont {L.}~\bibnamefont {Mai}}, \bibinfo {author}
  {\bibfnamefont {U.}~\bibnamefont {Scherf}},\ and\ \bibinfo {author}
  {\bibfnamefont {R.~F.}\ \bibnamefont {Mahrt}},\ }\bibfield  {title} {\bibinfo
  {title} {Room-temperature {Bose}–{Einstein} condensation of cavity
  exciton–polaritons in a polymer},\ }\href
  {https://doi.org/10.1038/nmat3825} {\bibfield  {journal} {\bibinfo  {journal}
  {Nature Mater}\ }\textbf {\bibinfo {volume} {13}},\ \bibinfo {pages} {247}
  (\bibinfo {year} {2014})}\BibitemShut {NoStop}%
\bibitem [{\citenamefont {De}\ \emph {et~al.}(2023)\citenamefont {De},
  \citenamefont {Ma}, \citenamefont {Yin}, \citenamefont {Ren}, \citenamefont
  {Yao}, \citenamefont {Schumacher}, \citenamefont {Liao}, \citenamefont {Fu},
  \citenamefont {Malpuech},\ and\ \citenamefont
  {Solnyshkov}}]{de_room-temperature_2023}%
  \BibitemOpen
  \bibfield  {author} {\bibinfo {author} {\bibfnamefont {J.}~\bibnamefont
  {De}}, \bibinfo {author} {\bibfnamefont {X.}~\bibnamefont {Ma}}, \bibinfo
  {author} {\bibfnamefont {F.}~\bibnamefont {Yin}}, \bibinfo {author}
  {\bibfnamefont {J.}~\bibnamefont {Ren}}, \bibinfo {author} {\bibfnamefont
  {J.}~\bibnamefont {Yao}}, \bibinfo {author} {\bibfnamefont {S.}~\bibnamefont
  {Schumacher}}, \bibinfo {author} {\bibfnamefont {Q.}~\bibnamefont {Liao}},
  \bibinfo {author} {\bibfnamefont {H.}~\bibnamefont {Fu}}, \bibinfo {author}
  {\bibfnamefont {G.}~\bibnamefont {Malpuech}},\ and\ \bibinfo {author}
  {\bibfnamefont {D.}~\bibnamefont {Solnyshkov}},\ }\bibfield  {title}
  {\bibinfo {title} {Room-{Temperature} {Electrical} {Field}-{Enhanced}
  {Ultrafast} {Switch} in {Organic} {Microcavity} {Polariton} {Condensates}},\
  }\href {https://doi.org/10.1021/jacs.2c07557} {\bibfield  {journal} {\bibinfo
   {journal} {J. Am. Chem. Soc.}\ }\textbf {\bibinfo {volume} {145}},\ \bibinfo
  {pages} {1557} (\bibinfo {year} {2023})}\BibitemShut {NoStop}%
\bibitem [{\citenamefont {Zasedatelev}\ \emph {et~al.}(2019)\citenamefont
  {Zasedatelev}, \citenamefont {Baranikov}, \citenamefont {Urbonas},
  \citenamefont {Scafirimuto}, \citenamefont {Scherf}, \citenamefont
  {Stöferle}, \citenamefont {Mahrt},\ and\ \citenamefont
  {Lagoudakis}}]{zasedatelev_room-temperature_2019}%
  \BibitemOpen
  \bibfield  {author} {\bibinfo {author} {\bibfnamefont {A.~V.}\ \bibnamefont
  {Zasedatelev}}, \bibinfo {author} {\bibfnamefont {A.~V.}\ \bibnamefont
  {Baranikov}}, \bibinfo {author} {\bibfnamefont {D.}~\bibnamefont {Urbonas}},
  \bibinfo {author} {\bibfnamefont {F.}~\bibnamefont {Scafirimuto}}, \bibinfo
  {author} {\bibfnamefont {U.}~\bibnamefont {Scherf}}, \bibinfo {author}
  {\bibfnamefont {T.}~\bibnamefont {Stöferle}}, \bibinfo {author}
  {\bibfnamefont {R.~F.}\ \bibnamefont {Mahrt}},\ and\ \bibinfo {author}
  {\bibfnamefont {P.~G.}\ \bibnamefont {Lagoudakis}},\ }\bibfield  {title}
  {\bibinfo {title} {A room-temperature organic polariton transistor},\ }\href
  {https://doi.org/10.1038/s41566-019-0392-8} {\bibfield  {journal} {\bibinfo
  {journal} {Nat. Photonics}\ }\textbf {\bibinfo {volume} {13}},\ \bibinfo
  {pages} {378} (\bibinfo {year} {2019})}\BibitemShut {NoStop}%
\bibitem [{\citenamefont {Laitz}\ \emph {et~al.}(2023)\citenamefont {Laitz},
  \citenamefont {Kaplan}, \citenamefont {Deschamps}, \citenamefont {Barotov},
  \citenamefont {Proppe}, \citenamefont {Garc{\'\i}a-Benito}, \citenamefont
  {Osherov}, \citenamefont {Grancini}, \citenamefont {deQuilettes},
  \citenamefont {Nelson} \emph {et~al.}}]{laitz2023uncovering}%
  \BibitemOpen
  \bibfield  {author} {\bibinfo {author} {\bibfnamefont {M.}~\bibnamefont
  {Laitz}}, \bibinfo {author} {\bibfnamefont {A.~E.}\ \bibnamefont {Kaplan}},
  \bibinfo {author} {\bibfnamefont {J.}~\bibnamefont {Deschamps}}, \bibinfo
  {author} {\bibfnamefont {U.}~\bibnamefont {Barotov}}, \bibinfo {author}
  {\bibfnamefont {A.~H.}\ \bibnamefont {Proppe}}, \bibinfo {author}
  {\bibfnamefont {I.}~\bibnamefont {Garc{\'\i}a-Benito}}, \bibinfo {author}
  {\bibfnamefont {A.}~\bibnamefont {Osherov}}, \bibinfo {author} {\bibfnamefont
  {G.}~\bibnamefont {Grancini}}, \bibinfo {author} {\bibfnamefont {D.~W.}\
  \bibnamefont {deQuilettes}}, \bibinfo {author} {\bibfnamefont {K.~A.}\
  \bibnamefont {Nelson}}, \emph {et~al.},\ }\bibfield  {title} {\bibinfo
  {title} {Uncovering temperature-dependent exciton-polariton relaxation
  mechanisms in hybrid organic-inorganic perovskites},\ }\href@noop {}
  {\bibfield  {journal} {\bibinfo  {journal} {Nature Communications}\ }\textbf
  {\bibinfo {volume} {14}},\ \bibinfo {pages} {2426} (\bibinfo {year}
  {2023})}\BibitemShut {NoStop}%
\bibitem [{\citenamefont {Xu}\ \emph {et~al.}(2023)\citenamefont {Xu},
  \citenamefont {Mandal}, \citenamefont {Baxter}, \citenamefont {Cheng},
  \citenamefont {Lee}, \citenamefont {Su}, \citenamefont {Liu}, \citenamefont
  {Reichman},\ and\ \citenamefont {Delor}}]{xu_ultrafast_2023}%
  \BibitemOpen
  \bibfield  {author} {\bibinfo {author} {\bibfnamefont {D.}~\bibnamefont
  {Xu}}, \bibinfo {author} {\bibfnamefont {A.}~\bibnamefont {Mandal}}, \bibinfo
  {author} {\bibfnamefont {J.~M.}\ \bibnamefont {Baxter}}, \bibinfo {author}
  {\bibfnamefont {S.-W.}\ \bibnamefont {Cheng}}, \bibinfo {author}
  {\bibfnamefont {I.}~\bibnamefont {Lee}}, \bibinfo {author} {\bibfnamefont
  {H.}~\bibnamefont {Su}}, \bibinfo {author} {\bibfnamefont {S.}~\bibnamefont
  {Liu}}, \bibinfo {author} {\bibfnamefont {D.~R.}\ \bibnamefont {Reichman}},\
  and\ \bibinfo {author} {\bibfnamefont {M.}~\bibnamefont {Delor}},\ }\bibfield
   {title} {\bibinfo {title} {Ultrafast imaging of polariton propagation and
  interactions},\ }\href {https://doi.org/10.1038/s41467-023-39550-x}
  {\bibfield  {journal} {\bibinfo  {journal} {Nat. Commun.}\ }\textbf {\bibinfo
  {volume} {14}},\ \bibinfo {pages} {3881} (\bibinfo {year}
  {2023})}\BibitemShut {NoStop}%
\bibitem [{\citenamefont {Balasubrahmaniyam}\ \emph {et~al.}(2023)\citenamefont
  {Balasubrahmaniyam}, \citenamefont {Simkhovich}, \citenamefont {Golombek},
  \citenamefont {Sandik}, \citenamefont {Ankonina},\ and\ \citenamefont
  {Schwartz}}]{balasubrahmaniyam_enhanced_2023}%
  \BibitemOpen
  \bibfield  {author} {\bibinfo {author} {\bibfnamefont {M.}~\bibnamefont
  {Balasubrahmaniyam}}, \bibinfo {author} {\bibfnamefont {A.}~\bibnamefont
  {Simkhovich}}, \bibinfo {author} {\bibfnamefont {A.}~\bibnamefont
  {Golombek}}, \bibinfo {author} {\bibfnamefont {G.}~\bibnamefont {Sandik}},
  \bibinfo {author} {\bibfnamefont {G.}~\bibnamefont {Ankonina}},\ and\
  \bibinfo {author} {\bibfnamefont {T.}~\bibnamefont {Schwartz}},\ }\bibfield
  {title} {\bibinfo {title} {From enhanced diffusion to ultrafast ballistic
  motion of hybrid light–matter excitations},\ }\href
  {https://doi.org/10.1038/s41563-022-01463-3} {\bibfield  {journal} {\bibinfo
  {journal} {Nat. Mater.}\ }\textbf {\bibinfo {volume} {22}},\ \bibinfo {pages}
  {338} (\bibinfo {year} {2023})}\BibitemShut {NoStop}%
\bibitem [{\citenamefont {Lerario}\ \emph {et~al.}(2017)\citenamefont
  {Lerario}, \citenamefont {Ballarini}, \citenamefont {Fieramosca},
  \citenamefont {Cannavale}, \citenamefont {Genco}, \citenamefont {Mangione},
  \citenamefont {Gambino}, \citenamefont {Dominici}, \citenamefont {De~Giorgi},
  \citenamefont {Gigli} \emph {et~al.}}]{lerario2017high}%
  \BibitemOpen
  \bibfield  {author} {\bibinfo {author} {\bibfnamefont {G.}~\bibnamefont
  {Lerario}}, \bibinfo {author} {\bibfnamefont {D.}~\bibnamefont {Ballarini}},
  \bibinfo {author} {\bibfnamefont {A.}~\bibnamefont {Fieramosca}}, \bibinfo
  {author} {\bibfnamefont {A.}~\bibnamefont {Cannavale}}, \bibinfo {author}
  {\bibfnamefont {A.}~\bibnamefont {Genco}}, \bibinfo {author} {\bibfnamefont
  {F.}~\bibnamefont {Mangione}}, \bibinfo {author} {\bibfnamefont
  {S.}~\bibnamefont {Gambino}}, \bibinfo {author} {\bibfnamefont
  {L.}~\bibnamefont {Dominici}}, \bibinfo {author} {\bibfnamefont
  {M.}~\bibnamefont {De~Giorgi}}, \bibinfo {author} {\bibfnamefont
  {G.}~\bibnamefont {Gigli}}, \emph {et~al.},\ }\bibfield  {title} {\bibinfo
  {title} {High-speed flow of interacting organic polaritons},\ }\href@noop {}
  {\bibfield  {journal} {\bibinfo  {journal} {Light: Science \& Applications}\
  }\textbf {\bibinfo {volume} {6}},\ \bibinfo {pages} {e16212} (\bibinfo {year}
  {2017})}\BibitemShut {NoStop}%
\bibitem [{\citenamefont {Myers}\ \emph {et~al.}(2018)\citenamefont {Myers},
  \citenamefont {Mukherjee}, \citenamefont {Beaumariage}, \citenamefont
  {Snoke}, \citenamefont {Steger}, \citenamefont {Pfeiffer},\ and\
  \citenamefont {West}}]{myers2018polariton}%
  \BibitemOpen
  \bibfield  {author} {\bibinfo {author} {\bibfnamefont {D.}~\bibnamefont
  {Myers}}, \bibinfo {author} {\bibfnamefont {S.}~\bibnamefont {Mukherjee}},
  \bibinfo {author} {\bibfnamefont {J.}~\bibnamefont {Beaumariage}}, \bibinfo
  {author} {\bibfnamefont {D.}~\bibnamefont {Snoke}}, \bibinfo {author}
  {\bibfnamefont {M.}~\bibnamefont {Steger}}, \bibinfo {author} {\bibfnamefont
  {L.}~\bibnamefont {Pfeiffer}},\ and\ \bibinfo {author} {\bibfnamefont
  {K.}~\bibnamefont {West}},\ }\bibfield  {title} {\bibinfo {title}
  {Polariton-enhanced exciton transport},\ }\href@noop {} {\bibfield  {journal}
  {\bibinfo  {journal} {Physical Review B}\ }\textbf {\bibinfo {volume} {98}},\
  \bibinfo {pages} {235302} (\bibinfo {year} {2018})}\BibitemShut {NoStop}%
\bibitem [{\citenamefont {Rozenman}\ \emph {et~al.}(2018)\citenamefont
  {Rozenman}, \citenamefont {Akulov}, \citenamefont {Golombek},\ and\
  \citenamefont {Schwartz}}]{rozenman2018long}%
  \BibitemOpen
  \bibfield  {author} {\bibinfo {author} {\bibfnamefont {G.~G.}\ \bibnamefont
  {Rozenman}}, \bibinfo {author} {\bibfnamefont {K.}~\bibnamefont {Akulov}},
  \bibinfo {author} {\bibfnamefont {A.}~\bibnamefont {Golombek}},\ and\
  \bibinfo {author} {\bibfnamefont {T.}~\bibnamefont {Schwartz}},\ }\bibfield
  {title} {\bibinfo {title} {Long-range transport of organic exciton-polaritons
  revealed by ultrafast microscopy},\ }\href@noop {} {\bibfield  {journal}
  {\bibinfo  {journal} {ACS photonics}\ }\textbf {\bibinfo {volume} {5}},\
  \bibinfo {pages} {105} (\bibinfo {year} {2018})}\BibitemShut {NoStop}%
\bibitem [{\citenamefont {Sokolovskii}\ \emph {et~al.}(2023)\citenamefont
  {Sokolovskii}, \citenamefont {Tichauer}, \citenamefont {Morozov},
  \citenamefont {Feist},\ and\ \citenamefont
  {Groenhof}}]{sokolovskii_multi-scale_2023}%
  \BibitemOpen
  \bibfield  {author} {\bibinfo {author} {\bibfnamefont {I.}~\bibnamefont
  {Sokolovskii}}, \bibinfo {author} {\bibfnamefont {R.~H.}\ \bibnamefont
  {Tichauer}}, \bibinfo {author} {\bibfnamefont {D.}~\bibnamefont {Morozov}},
  \bibinfo {author} {\bibfnamefont {J.}~\bibnamefont {Feist}},\ and\ \bibinfo
  {author} {\bibfnamefont {G.}~\bibnamefont {Groenhof}},\ }\bibfield  {title}
  {\bibinfo {title} {Multi-scale molecular dynamics simulations of enhanced
  energy transfer in organic molecules under strong coupling},\ }\href
  {https://doi.org/10.1038/s41467-023-42067-y} {\bibfield  {journal} {\bibinfo
  {journal} {Nat. Commun.}\ }\textbf {\bibinfo {volume} {14}},\ \bibinfo
  {pages} {6613} (\bibinfo {year} {2023})}\BibitemShut {NoStop}%
\bibitem [{\citenamefont {Sanvitto}\ and\ \citenamefont
  {Kéna-Cohen}(2016)}]{sanvitto_road_2016}%
  \BibitemOpen
  \bibfield  {author} {\bibinfo {author} {\bibfnamefont {D.}~\bibnamefont
  {Sanvitto}}\ and\ \bibinfo {author} {\bibfnamefont {S.}~\bibnamefont
  {Kéna-Cohen}},\ }\bibfield  {title} {\bibinfo {title} {The road towards
  polaritonic devices},\ }\href {https://doi.org/10.1038/nmat4668} {\bibfield
  {journal} {\bibinfo  {journal} {Nature Mater}\ }\textbf {\bibinfo {volume}
  {15}},\ \bibinfo {pages} {1061} (\bibinfo {year} {2016})}\BibitemShut
  {NoStop}%
\bibitem [{\citenamefont {Rafique}\ \emph {et~al.}(2018)\citenamefont
  {Rafique}, \citenamefont {Abdullah}, \citenamefont {Sulaiman},\ and\
  \citenamefont {Iwamoto}}]{rafique_fundamentals_2018}%
  \BibitemOpen
  \bibfield  {author} {\bibinfo {author} {\bibfnamefont {S.}~\bibnamefont
  {Rafique}}, \bibinfo {author} {\bibfnamefont {S.~M.}\ \bibnamefont
  {Abdullah}}, \bibinfo {author} {\bibfnamefont {K.}~\bibnamefont {Sulaiman}},\
  and\ \bibinfo {author} {\bibfnamefont {M.}~\bibnamefont {Iwamoto}},\
  }\bibfield  {title} {\bibinfo {title} {Fundamentals of bulk heterojunction
  organic solar cells: {An} overview of stability/degradation issues and
  strategies for improvement},\ }\href
  {https://doi.org/10.1016/j.rser.2017.12.008} {\bibfield  {journal} {\bibinfo
  {journal} {Ren. Sust. Energ. Rev.}\ }\textbf {\bibinfo {volume} {84}},\
  \bibinfo {pages} {43} (\bibinfo {year} {2018})}\BibitemShut {NoStop}%
\bibitem [{\citenamefont {Freixanet}\ \emph {et~al.}(2000)\citenamefont
  {Freixanet}, \citenamefont {Sermage}, \citenamefont {Tiberj},\ and\
  \citenamefont {Planel}}]{freixanet_-plane_2000}%
  \BibitemOpen
  \bibfield  {author} {\bibinfo {author} {\bibfnamefont {T.}~\bibnamefont
  {Freixanet}}, \bibinfo {author} {\bibfnamefont {B.}~\bibnamefont {Sermage}},
  \bibinfo {author} {\bibfnamefont {A.}~\bibnamefont {Tiberj}},\ and\ \bibinfo
  {author} {\bibfnamefont {R.}~\bibnamefont {Planel}},\ }\bibfield  {title}
  {\bibinfo {title} {In-plane propagation of excitonic cavity polaritons},\
  }\href {https://doi.org/10.1103/PhysRevB.61.7233} {\bibfield  {journal}
  {\bibinfo  {journal} {Phys. Rev. B}\ }\textbf {\bibinfo {volume} {61}},\
  \bibinfo {pages} {7233} (\bibinfo {year} {2000})}\BibitemShut {NoStop}%
\bibitem [{\citenamefont {Coles}\ \emph {et~al.}(2011)\citenamefont {Coles},
  \citenamefont {Michetti}, \citenamefont {Clark}, \citenamefont {Tsoi},
  \citenamefont {Adawi}, \citenamefont {Kim},\ and\ \citenamefont
  {Lidzey}}]{coles2011vibrationally}%
  \BibitemOpen
  \bibfield  {author} {\bibinfo {author} {\bibfnamefont {D.~M.}\ \bibnamefont
  {Coles}}, \bibinfo {author} {\bibfnamefont {P.}~\bibnamefont {Michetti}},
  \bibinfo {author} {\bibfnamefont {C.}~\bibnamefont {Clark}}, \bibinfo
  {author} {\bibfnamefont {W.~C.}\ \bibnamefont {Tsoi}}, \bibinfo {author}
  {\bibfnamefont {A.~M.}\ \bibnamefont {Adawi}}, \bibinfo {author}
  {\bibfnamefont {J.-S.}\ \bibnamefont {Kim}},\ and\ \bibinfo {author}
  {\bibfnamefont {D.~G.}\ \bibnamefont {Lidzey}},\ }\bibfield  {title}
  {\bibinfo {title} {Vibrationally assisted polariton-relaxation processes in
  strongly coupled organic-semiconductor microcavities},\ }\href@noop {}
  {\bibfield  {journal} {\bibinfo  {journal} {Advanced Functional Materials}\
  }\textbf {\bibinfo {volume} {21}},\ \bibinfo {pages} {3691} (\bibinfo {year}
  {2011})}\BibitemShut {NoStop}%
\bibitem [{\citenamefont {Agranovich}\ \emph {et~al.}(2003)\citenamefont
  {Agranovich}, \citenamefont {Litinskaia},\ and\ \citenamefont
  {Lidzey}}]{agranovich_cavity_2003}%
  \BibitemOpen
  \bibfield  {author} {\bibinfo {author} {\bibfnamefont {V.~M.}\ \bibnamefont
  {Agranovich}}, \bibinfo {author} {\bibfnamefont {M.}~\bibnamefont
  {Litinskaia}},\ and\ \bibinfo {author} {\bibfnamefont {D.~G.}\ \bibnamefont
  {Lidzey}},\ }\bibfield  {title} {\bibinfo {title} {Cavity polaritons in
  microcavities containing disordered organic semiconductors},\ }\href
  {https://doi.org/10.1103/PhysRevB.67.085311} {\bibfield  {journal} {\bibinfo
  {journal} {Phys. Rev. B}\ }\textbf {\bibinfo {volume} {67}},\ \bibinfo
  {pages} {085311} (\bibinfo {year} {2003})}\BibitemShut {NoStop}%
\bibitem [{\citenamefont {Coles}\ \emph {et~al.}(2013)\citenamefont {Coles},
  \citenamefont {Grant}, \citenamefont {Lidzey}, \citenamefont {Clark},\ and\
  \citenamefont {Lagoudakis}}]{coles2013imaging}%
  \BibitemOpen
  \bibfield  {author} {\bibinfo {author} {\bibfnamefont {D.~M.}\ \bibnamefont
  {Coles}}, \bibinfo {author} {\bibfnamefont {R.~T.}\ \bibnamefont {Grant}},
  \bibinfo {author} {\bibfnamefont {D.~G.}\ \bibnamefont {Lidzey}}, \bibinfo
  {author} {\bibfnamefont {C.}~\bibnamefont {Clark}},\ and\ \bibinfo {author}
  {\bibfnamefont {P.~G.}\ \bibnamefont {Lagoudakis}},\ }\bibfield  {title}
  {\bibinfo {title} {Imaging the polariton relaxation bottleneck in strongly
  coupled organic semiconductor microcavities},\ }\href@noop {} {\bibfield
  {journal} {\bibinfo  {journal} {Physical Review B}\ }\textbf {\bibinfo
  {volume} {88}},\ \bibinfo {pages} {121303} (\bibinfo {year}
  {2013})}\BibitemShut {NoStop}%
\bibitem [{\citenamefont {Somaschi}\ \emph {et~al.}(2011)\citenamefont
  {Somaschi}, \citenamefont {Mouchliadis}, \citenamefont {Coles}, \citenamefont
  {Perakis}, \citenamefont {Lidzey}, \citenamefont {Lagoudakis},\ and\
  \citenamefont {Savvidis}}]{somaschi2011ultrafast}%
  \BibitemOpen
  \bibfield  {author} {\bibinfo {author} {\bibfnamefont {N.}~\bibnamefont
  {Somaschi}}, \bibinfo {author} {\bibfnamefont {L.}~\bibnamefont
  {Mouchliadis}}, \bibinfo {author} {\bibfnamefont {D.}~\bibnamefont {Coles}},
  \bibinfo {author} {\bibfnamefont {I.}~\bibnamefont {Perakis}}, \bibinfo
  {author} {\bibfnamefont {D.}~\bibnamefont {Lidzey}}, \bibinfo {author}
  {\bibfnamefont {P.}~\bibnamefont {Lagoudakis}},\ and\ \bibinfo {author}
  {\bibfnamefont {P.}~\bibnamefont {Savvidis}},\ }\bibfield  {title} {\bibinfo
  {title} {Ultrafast polariton population build-up mediated by molecular
  phonons in organic microcavities},\ }\href@noop {} {\bibfield  {journal}
  {\bibinfo  {journal} {Applied Physics Letters}\ }\textbf {\bibinfo {volume}
  {99}} (\bibinfo {year} {2011})}\BibitemShut {NoStop}%
\bibitem [{\citenamefont {Lüttgens}\ \emph {et~al.}(2020)\citenamefont
  {Lüttgens}, \citenamefont {Berger},\ and\ \citenamefont
  {Zaumseil}}]{luttgens2020population}%
  \BibitemOpen
  \bibfield  {author} {\bibinfo {author} {\bibfnamefont {J.~M.}\ \bibnamefont
  {Lüttgens}}, \bibinfo {author} {\bibfnamefont {F.~J.}\ \bibnamefont
  {Berger}},\ and\ \bibinfo {author} {\bibfnamefont {J.}~\bibnamefont
  {Zaumseil}},\ }\bibfield  {title} {\bibinfo {title} {Population of
  exciton--polaritons via luminescent sp3 defects in single-walled carbon
  nanotubes},\ }\href@noop {} {\bibfield  {journal} {\bibinfo  {journal} {ACS
  photonics}\ }\textbf {\bibinfo {volume} {8}},\ \bibinfo {pages} {182}
  (\bibinfo {year} {2020})}\BibitemShut {NoStop}%
\bibitem [{\citenamefont {Virgili}\ \emph {et~al.}(2011)\citenamefont
  {Virgili}, \citenamefont {Coles}, \citenamefont {Adawi}, \citenamefont
  {Clark}, \citenamefont {Michetti}, \citenamefont {Rajendran}, \citenamefont
  {Brida}, \citenamefont {Polli}, \citenamefont {Cerullo},\ and\ \citenamefont
  {Lidzey}}]{virgili_ultrafast_2011}%
  \BibitemOpen
  \bibfield  {author} {\bibinfo {author} {\bibfnamefont {T.}~\bibnamefont
  {Virgili}}, \bibinfo {author} {\bibfnamefont {D.}~\bibnamefont {Coles}},
  \bibinfo {author} {\bibfnamefont {A.~M.}\ \bibnamefont {Adawi}}, \bibinfo
  {author} {\bibfnamefont {C.}~\bibnamefont {Clark}}, \bibinfo {author}
  {\bibfnamefont {P.}~\bibnamefont {Michetti}}, \bibinfo {author}
  {\bibfnamefont {S.~K.}\ \bibnamefont {Rajendran}}, \bibinfo {author}
  {\bibfnamefont {D.}~\bibnamefont {Brida}}, \bibinfo {author} {\bibfnamefont
  {D.}~\bibnamefont {Polli}}, \bibinfo {author} {\bibfnamefont
  {G.}~\bibnamefont {Cerullo}},\ and\ \bibinfo {author} {\bibfnamefont {D.~G.}\
  \bibnamefont {Lidzey}},\ }\bibfield  {title} {\bibinfo {title} {Ultrafast
  polariton relaxation dynamics in an organic semiconductor microcavity},\
  }\href {https://doi.org/10.1103/PhysRevB.83.245309} {\bibfield  {journal}
  {\bibinfo  {journal} {Phys. Rev. B}\ }\textbf {\bibinfo {volume} {83}},\
  \bibinfo {pages} {245309} (\bibinfo {year} {2011})}\BibitemShut {NoStop}%
\bibitem [{\citenamefont {Pandya}\ \emph {et~al.}(2021)\citenamefont {Pandya},
  \citenamefont {Chen}, \citenamefont {Gu}, \citenamefont {Sung}, \citenamefont
  {Schnedermann}, \citenamefont {Ojambati}, \citenamefont {Chikkaraddy},
  \citenamefont {Gorman}, \citenamefont {Jacucci}, \citenamefont {Onelli},
  \citenamefont {Willhammar}, \citenamefont {Johnstone}, \citenamefont
  {Collins}, \citenamefont {Midgley}, \citenamefont {Auras}, \citenamefont
  {Baikie}, \citenamefont {Jayaprakash}, \citenamefont {Mathevet},
  \citenamefont {Soucek}, \citenamefont {Du}, \citenamefont {Alvertis},
  \citenamefont {Ashoka}, \citenamefont {Vignolini}, \citenamefont {Lidzey},
  \citenamefont {Baumberg}, \citenamefont {Friend}, \citenamefont {Barisien},
  \citenamefont {Legrand}, \citenamefont {Chin}, \citenamefont {Yuen-Zhou},
  \citenamefont {Saikin}, \citenamefont {Kukura}, \citenamefont {Musser},\ and\
  \citenamefont {Rao}}]{pandya_microcavity-like_2021}%
  \BibitemOpen
  \bibfield  {author} {\bibinfo {author} {\bibfnamefont {R.}~\bibnamefont
  {Pandya}}, \bibinfo {author} {\bibfnamefont {R.~Y.~S.}\ \bibnamefont {Chen}},
  \bibinfo {author} {\bibfnamefont {Q.}~\bibnamefont {Gu}}, \bibinfo {author}
  {\bibfnamefont {J.}~\bibnamefont {Sung}}, \bibinfo {author} {\bibfnamefont
  {C.}~\bibnamefont {Schnedermann}}, \bibinfo {author} {\bibfnamefont {O.~S.}\
  \bibnamefont {Ojambati}}, \bibinfo {author} {\bibfnamefont {R.}~\bibnamefont
  {Chikkaraddy}}, \bibinfo {author} {\bibfnamefont {J.}~\bibnamefont {Gorman}},
  \bibinfo {author} {\bibfnamefont {G.}~\bibnamefont {Jacucci}}, \bibinfo
  {author} {\bibfnamefont {O.~D.}\ \bibnamefont {Onelli}}, \bibinfo {author}
  {\bibfnamefont {T.}~\bibnamefont {Willhammar}}, \bibinfo {author}
  {\bibfnamefont {D.~N.}\ \bibnamefont {Johnstone}}, \bibinfo {author}
  {\bibfnamefont {S.~M.}\ \bibnamefont {Collins}}, \bibinfo {author}
  {\bibfnamefont {P.~A.}\ \bibnamefont {Midgley}}, \bibinfo {author}
  {\bibfnamefont {F.}~\bibnamefont {Auras}}, \bibinfo {author} {\bibfnamefont
  {T.}~\bibnamefont {Baikie}}, \bibinfo {author} {\bibfnamefont
  {R.}~\bibnamefont {Jayaprakash}}, \bibinfo {author} {\bibfnamefont
  {F.}~\bibnamefont {Mathevet}}, \bibinfo {author} {\bibfnamefont
  {R.}~\bibnamefont {Soucek}}, \bibinfo {author} {\bibfnamefont
  {M.}~\bibnamefont {Du}}, \bibinfo {author} {\bibfnamefont {A.~M.}\
  \bibnamefont {Alvertis}}, \bibinfo {author} {\bibfnamefont {A.}~\bibnamefont
  {Ashoka}}, \bibinfo {author} {\bibfnamefont {S.}~\bibnamefont {Vignolini}},
  \bibinfo {author} {\bibfnamefont {D.~G.}\ \bibnamefont {Lidzey}}, \bibinfo
  {author} {\bibfnamefont {J.~J.}\ \bibnamefont {Baumberg}}, \bibinfo {author}
  {\bibfnamefont {R.~H.}\ \bibnamefont {Friend}}, \bibinfo {author}
  {\bibfnamefont {T.}~\bibnamefont {Barisien}}, \bibinfo {author}
  {\bibfnamefont {L.}~\bibnamefont {Legrand}}, \bibinfo {author} {\bibfnamefont
  {A.~W.}\ \bibnamefont {Chin}}, \bibinfo {author} {\bibfnamefont
  {J.}~\bibnamefont {Yuen-Zhou}}, \bibinfo {author} {\bibfnamefont {S.~K.}\
  \bibnamefont {Saikin}}, \bibinfo {author} {\bibfnamefont {P.}~\bibnamefont
  {Kukura}}, \bibinfo {author} {\bibfnamefont {A.~J.}\ \bibnamefont {Musser}},\
  and\ \bibinfo {author} {\bibfnamefont {A.}~\bibnamefont {Rao}},\ }\bibfield
  {title} {\bibinfo {title} {Microcavity-like exciton-polaritons can be the
  primary photoexcitation in bare organic semiconductors},\ }\href
  {https://doi.org/10.1038/s41467-021-26617-w} {\bibfield  {journal} {\bibinfo
  {journal} {Nat. Commun.}\ }\textbf {\bibinfo {volume} {12}},\ \bibinfo
  {pages} {6519} (\bibinfo {year} {2021})}\BibitemShut {NoStop}%
\bibitem [{\citenamefont {Sokolovskii}\ and\ \citenamefont
  {Groenhof}(2024)}]{sokolovskii_photochemical_2024}%
  \BibitemOpen
  \bibfield  {author} {\bibinfo {author} {\bibfnamefont {I.}~\bibnamefont
  {Sokolovskii}}\ and\ \bibinfo {author} {\bibfnamefont {G.}~\bibnamefont
  {Groenhof}},\ }\bibfield  {title} {\bibinfo {title} {Photochemical initiation
  of polariton-mediated exciton propagation},\ }\href@noop {} {\bibfield
  {journal} {\bibinfo  {journal} {Nanophotonics}\ }\textbf {\bibinfo {volume}
  {13}},\ \bibinfo {pages} {2687} (\bibinfo {year} {2024})}\BibitemShut
  {NoStop}%
\bibitem [{\citenamefont {Tichauer}\ \emph {et~al.}(2023)\citenamefont
  {Tichauer}, \citenamefont {Sokolovskii},\ and\ \citenamefont
  {Groenhof}}]{tichauer_tuning_2023}%
  \BibitemOpen
  \bibfield  {author} {\bibinfo {author} {\bibfnamefont {R.~H.}\ \bibnamefont
  {Tichauer}}, \bibinfo {author} {\bibfnamefont {I.}~\bibnamefont
  {Sokolovskii}},\ and\ \bibinfo {author} {\bibfnamefont {G.}~\bibnamefont
  {Groenhof}},\ }\bibfield  {title} {\bibinfo {title} {Tuning the {Coherent}
  {Propagation} of {Organic} {Exciton}‐{Polaritons} through the {Cavity}
  {Q}‐factor},\ }\href {https://doi.org/10.1002/advs.202302650} {\bibfield
  {journal} {\bibinfo  {journal} {Adv. Sci.}\ }\textbf {\bibinfo {volume}
  {10}},\ \bibinfo {pages} {2302650} (\bibinfo {year} {2023})}\BibitemShut
  {NoStop}%
\bibitem [{\citenamefont {Wang}\ and\ \citenamefont
  {Thoss}(2003)}]{wan_03_1289}%
  \BibitemOpen
  \bibfield  {author} {\bibinfo {author} {\bibfnamefont {H.}~\bibnamefont
  {Wang}}\ and\ \bibinfo {author} {\bibfnamefont {M.}~\bibnamefont {Thoss}},\
  }\bibfield  {title} {\bibinfo {title} {Multilayer formulation of the
  multiconfiguration time-dependent {{Hartree}} theory},\ }\href@noop {}
  {\bibfield  {journal} {\bibinfo  {journal} {J. Chem. Phys.}\ }\textbf
  {\bibinfo {volume} {119}},\ \bibinfo {pages} {1289} (\bibinfo {year}
  {2003})}\BibitemShut {NoStop}%
\bibitem [{\citenamefont {Manthe}(2008)}]{man_08_164116}%
  \BibitemOpen
  \bibfield  {author} {\bibinfo {author} {\bibfnamefont {U.}~\bibnamefont
  {Manthe}},\ }\bibfield  {title} {\bibinfo {title} {A multilayer
  multiconfigurational time-dependent {{Hartree}} approach for quantum dynamics
  on general potential energy surfaces},\ }\href
  {https://doi.org/10.1063/1.2902982} {\bibfield  {journal} {\bibinfo
  {journal} {J. Chem. Phys.}\ }\textbf {\bibinfo {volume} {128}},\ \bibinfo
  {pages} {164116} (\bibinfo {year} {2008})}\BibitemShut {NoStop}%
\bibitem [{\citenamefont {Vendrell}\ and\ \citenamefont
  {Meyer}(2011)}]{vendrell_multilayer_2011}%
  \BibitemOpen
  \bibfield  {author} {\bibinfo {author} {\bibfnamefont {O.}~\bibnamefont
  {Vendrell}}\ and\ \bibinfo {author} {\bibfnamefont {H.-D.}\ \bibnamefont
  {Meyer}},\ }\bibfield  {title} {\bibinfo {title} {Multilayer
  multiconfiguration time-dependent {Hartree} method: {Implementation} and
  applications to a {Henon}–{Heiles} {Hamiltonian} and to pyrazine},\ }\href
  {https://doi.org/10.1063/1.3535541} {\bibfield  {journal} {\bibinfo
  {journal} {J. Chem. Phys.}\ }\textbf {\bibinfo {volume} {134}},\ \bibinfo
  {pages} {044135} (\bibinfo {year} {2011})}\BibitemShut {NoStop}%
\bibitem [{\citenamefont
  {Vendrell}(2018{\natexlab{a}})}]{vendrell_coherent_2018}%
  \BibitemOpen
  \bibfield  {author} {\bibinfo {author} {\bibfnamefont {O.}~\bibnamefont
  {Vendrell}},\ }\bibfield  {title} {\bibinfo {title} {Coherent dynamics in
  cavity femtochemistry: {Application} of the multi-configuration
  time-dependent {Hartree} method},\ }\href
  {https://doi.org/10.1016/j.chemphys.2018.02.008} {\bibfield  {journal}
  {\bibinfo  {journal} {Chem. Phys.}\ }\textbf {\bibinfo {volume} {509}},\
  \bibinfo {pages} {55} (\bibinfo {year} {2018}{\natexlab{a}})}\BibitemShut
  {NoStop}%
\bibitem [{\citenamefont {Michetti}\ and\ \citenamefont
  {Rocca}(2005)}]{Michetti2005}%
  \BibitemOpen
  \bibfield  {author} {\bibinfo {author} {\bibfnamefont {P.}~\bibnamefont
  {Michetti}}\ and\ \bibinfo {author} {\bibfnamefont {G.~C.~L.}\ \bibnamefont
  {Rocca}},\ }\bibfield  {title} {\bibinfo {title} {Polariton states in
  disordered organic microcavities},\ }\href@noop {} {\bibfield  {journal}
  {\bibinfo  {journal} {Phys. Rev. B.}\ }\textbf {\bibinfo {volume} {71}},\
  \bibinfo {pages} {115320} (\bibinfo {year} {2005})}\BibitemShut {NoStop}%
\bibitem [{\citenamefont {Worth}\ \emph {et~al.}(1996)\citenamefont {Worth},
  \citenamefont {Meyer},\ and\ \citenamefont {Cederbaum}}]{worth1996effect}%
  \BibitemOpen
  \bibfield  {author} {\bibinfo {author} {\bibfnamefont {G.~A.}\ \bibnamefont
  {Worth}}, \bibinfo {author} {\bibfnamefont {H.-D.}\ \bibnamefont {Meyer}},\
  and\ \bibinfo {author} {\bibfnamefont {L.}~\bibnamefont {Cederbaum}},\
  }\bibfield  {title} {\bibinfo {title} {The effect of a model environment on
  the s 2 absorption spectrum of pyrazine: A wave packet study treating all 24
  vibrational modes},\ }\href@noop {} {\bibfield  {journal} {\bibinfo
  {journal} {The Journal of chemical physics}\ }\textbf {\bibinfo {volume}
  {105}},\ \bibinfo {pages} {4412} (\bibinfo {year} {1996})}\BibitemShut
  {NoStop}%
\bibitem [{\citenamefont {Terry~Weatherly}\ \emph {et~al.}(2023)\citenamefont
  {Terry~Weatherly}, \citenamefont {Provazza}, \citenamefont {Weiss},\ and\
  \citenamefont {Tempelaar}}]{terry2023theory}%
  \BibitemOpen
  \bibfield  {author} {\bibinfo {author} {\bibfnamefont {C.~K.}\ \bibnamefont
  {Terry~Weatherly}}, \bibinfo {author} {\bibfnamefont {J.}~\bibnamefont
  {Provazza}}, \bibinfo {author} {\bibfnamefont {E.~A.}\ \bibnamefont
  {Weiss}},\ and\ \bibinfo {author} {\bibfnamefont {R.}~\bibnamefont
  {Tempelaar}},\ }\bibfield  {title} {\bibinfo {title} {Theory predicts
  uv/vis-to-ir photonic down conversion mediated by excited state vibrational
  polaritons},\ }\href@noop {} {\bibfield  {journal} {\bibinfo  {journal}
  {Nature communications}\ }\textbf {\bibinfo {volume} {14}},\ \bibinfo {pages}
  {4804} (\bibinfo {year} {2023})}\BibitemShut {NoStop}%
\bibitem [{\citenamefont {Barclay}\ \emph {et~al.}(2022)\citenamefont
  {Barclay}, \citenamefont {Huff}, \citenamefont {Pensack}, \citenamefont
  {Davis}, \citenamefont {Knowlton}, \citenamefont {Yurke}, \citenamefont
  {Dean}, \citenamefont {Arpin},\ and\ \citenamefont
  {Turner}}]{barclay2022characterizing}%
  \BibitemOpen
  \bibfield  {author} {\bibinfo {author} {\bibfnamefont {M.~S.}\ \bibnamefont
  {Barclay}}, \bibinfo {author} {\bibfnamefont {J.~S.}\ \bibnamefont {Huff}},
  \bibinfo {author} {\bibfnamefont {R.~D.}\ \bibnamefont {Pensack}}, \bibinfo
  {author} {\bibfnamefont {P.~H.}\ \bibnamefont {Davis}}, \bibinfo {author}
  {\bibfnamefont {W.~B.}\ \bibnamefont {Knowlton}}, \bibinfo {author}
  {\bibfnamefont {B.}~\bibnamefont {Yurke}}, \bibinfo {author} {\bibfnamefont
  {J.~C.}\ \bibnamefont {Dean}}, \bibinfo {author} {\bibfnamefont {P.~C.}\
  \bibnamefont {Arpin}},\ and\ \bibinfo {author} {\bibfnamefont {D.~B.}\
  \bibnamefont {Turner}},\ }\bibfield  {title} {\bibinfo {title}
  {Characterizing mode anharmonicity and huang--rhys factors using models of
  femtosecond coherence spectra},\ }\href@noop {} {\bibfield  {journal}
  {\bibinfo  {journal} {The Journal of Physical Chemistry Letters}\ }\textbf
  {\bibinfo {volume} {13}},\ \bibinfo {pages} {5413} (\bibinfo {year}
  {2022})}\BibitemShut {NoStop}%
\bibitem [{\citenamefont {Ćwik}\ \emph {et~al.}(2016)\citenamefont {Ćwik},
  \citenamefont {Kirton}, \citenamefont {De~Liberato},\ and\ \citenamefont
  {Keeling}}]{cwik_excitonic_2016}%
  \BibitemOpen
  \bibfield  {author} {\bibinfo {author} {\bibfnamefont {J.~A.}\ \bibnamefont
  {Ćwik}}, \bibinfo {author} {\bibfnamefont {P.}~\bibnamefont {Kirton}},
  \bibinfo {author} {\bibfnamefont {S.}~\bibnamefont {De~Liberato}},\ and\
  \bibinfo {author} {\bibfnamefont {J.}~\bibnamefont {Keeling}},\ }\bibfield
  {title} {\bibinfo {title} {Excitonic spectral features in strongly coupled
  organic polaritons},\ }\href {https://doi.org/10.1103/PhysRevA.93.033840}
  {\bibfield  {journal} {\bibinfo  {journal} {Phys. Rev. A}\ }\textbf {\bibinfo
  {volume} {93}},\ \bibinfo {pages} {033840} (\bibinfo {year}
  {2016})}\BibitemShut {NoStop}%
\bibitem [{\citenamefont {Pandya}\ \emph {et~al.}(2022)\citenamefont {Pandya},
  \citenamefont {Ashoka}, \citenamefont {Georgiou}, \citenamefont {Sung},
  \citenamefont {Jayaprakash}, \citenamefont {Renken}, \citenamefont {Gai},
  \citenamefont {Shen}, \citenamefont {Rao},\ and\ \citenamefont
  {Musser}}]{pandya2022tuning}%
  \BibitemOpen
  \bibfield  {author} {\bibinfo {author} {\bibfnamefont {R.}~\bibnamefont
  {Pandya}}, \bibinfo {author} {\bibfnamefont {A.}~\bibnamefont {Ashoka}},
  \bibinfo {author} {\bibfnamefont {K.}~\bibnamefont {Georgiou}}, \bibinfo
  {author} {\bibfnamefont {J.}~\bibnamefont {Sung}}, \bibinfo {author}
  {\bibfnamefont {R.}~\bibnamefont {Jayaprakash}}, \bibinfo {author}
  {\bibfnamefont {S.}~\bibnamefont {Renken}}, \bibinfo {author} {\bibfnamefont
  {L.}~\bibnamefont {Gai}}, \bibinfo {author} {\bibfnamefont {Z.}~\bibnamefont
  {Shen}}, \bibinfo {author} {\bibfnamefont {A.}~\bibnamefont {Rao}},\ and\
  \bibinfo {author} {\bibfnamefont {A.~J.}\ \bibnamefont {Musser}},\ }\bibfield
   {title} {\bibinfo {title} {Tuning the coherent propagation of organic
  exciton-polaritons through dark state delocalization},\ }\href@noop {}
  {\bibfield  {journal} {\bibinfo  {journal} {Advanced Science}\ }\textbf
  {\bibinfo {volume} {9}},\ \bibinfo {pages} {2105569} (\bibinfo {year}
  {2022})}\BibitemShut {NoStop}%
\bibitem [{\citenamefont {Litinskaya}\ \emph {et~al.}(2004)\citenamefont
  {Litinskaya}, \citenamefont {Reineker},\ and\ \citenamefont
  {Agranovich}}]{litinskaya_fast_2004}%
  \BibitemOpen
  \bibfield  {author} {\bibinfo {author} {\bibfnamefont {M.}~\bibnamefont
  {Litinskaya}}, \bibinfo {author} {\bibfnamefont {P.}~\bibnamefont
  {Reineker}},\ and\ \bibinfo {author} {\bibfnamefont {V.}~\bibnamefont
  {Agranovich}},\ }\bibfield  {title} {\bibinfo {title} {Fast polariton
  relaxation in strongly coupled organic microcavities},\ }\href
  {https://doi.org/10.1016/j.jlumin.2004.08.033} {\bibfield  {journal}
  {\bibinfo  {journal} {J. Lumin.}\ }\textbf {\bibinfo {volume} {110}},\
  \bibinfo {pages} {364} (\bibinfo {year} {2004})}\BibitemShut {NoStop}%
\bibitem [{\citenamefont {Tichauer}\ \emph {et~al.}(2022)\citenamefont
  {Tichauer}, \citenamefont {Morozov}, \citenamefont {Sokolovskii},
  \citenamefont {Toppari},\ and\ \citenamefont
  {Groenhof}}]{tichauer_identifying_2022}%
  \BibitemOpen
  \bibfield  {author} {\bibinfo {author} {\bibfnamefont {R.~H.}\ \bibnamefont
  {Tichauer}}, \bibinfo {author} {\bibfnamefont {D.}~\bibnamefont {Morozov}},
  \bibinfo {author} {\bibfnamefont {I.}~\bibnamefont {Sokolovskii}}, \bibinfo
  {author} {\bibfnamefont {J.~J.}\ \bibnamefont {Toppari}},\ and\ \bibinfo
  {author} {\bibfnamefont {G.}~\bibnamefont {Groenhof}},\ }\bibfield  {title}
  {\bibinfo {title} {Identifying {Vibrations} that {Control} {Non}-adiabatic
  {Relaxation} of {Polaritons} in {Strongly} {Coupled} {Molecule}–{Cavity}
  {Systems}},\ }\href {https://doi.org/10.1021/acs.jpclett.2c00826} {\bibfield
  {journal} {\bibinfo  {journal} {J. Phys. Chem. Lett.}\ }\textbf {\bibinfo
  {volume} {13}},\ \bibinfo {pages} {6259} (\bibinfo {year}
  {2022})}\BibitemShut {NoStop}%
\bibitem [{\citenamefont {Keeling}\ and\ \citenamefont
  {Kéna-Cohen}(2020)}]{keeling_boseeinstein_2020}%
  \BibitemOpen
  \bibfield  {author} {\bibinfo {author} {\bibfnamefont {J.}~\bibnamefont
  {Keeling}}\ and\ \bibinfo {author} {\bibfnamefont {S.}~\bibnamefont
  {Kéna-Cohen}},\ }\bibfield  {title} {\bibinfo {title} {Bose–{Einstein}
  {Condensation} of {Exciton}-{Polaritons} in {Organic} {Microcavities}},\
  }\href {https://doi.org/10.1146/annurev-physchem-010920-102509} {\bibfield
  {journal} {\bibinfo  {journal} {Annu. Rev. Phys. Chem.}\ }\textbf {\bibinfo
  {volume} {71}},\ \bibinfo {pages} {435} (\bibinfo {year} {2020})}\BibitemShut
  {NoStop}%
\bibitem [{\citenamefont {Agranovich}\ and\ \citenamefont
  {Gartstein}(2007)}]{agranovich_nature_2007}%
  \BibitemOpen
  \bibfield  {author} {\bibinfo {author} {\bibfnamefont {V.~M.}\ \bibnamefont
  {Agranovich}}\ and\ \bibinfo {author} {\bibfnamefont {Y.~N.}\ \bibnamefont
  {Gartstein}},\ }\bibfield  {title} {\bibinfo {title} {Nature and dynamics of
  low-energy exciton polaritons in semiconductor microcavities},\ }\href
  {https://doi.org/10.1103/PhysRevB.75.075302} {\bibfield  {journal} {\bibinfo
  {journal} {Phys. Rev. B}\ }\textbf {\bibinfo {volume} {75}},\ \bibinfo
  {pages} {075302} (\bibinfo {year} {2007})}\BibitemShut {NoStop}%
\bibitem [{\citenamefont
  {Vendrell}(2018{\natexlab{b}})}]{vendrell_collective_2018}%
  \BibitemOpen
  \bibfield  {author} {\bibinfo {author} {\bibfnamefont {O.}~\bibnamefont
  {Vendrell}},\ }\bibfield  {title} {\bibinfo {title} {Collective
  {Jahn}-{Teller} {Interactions} through {Light}-{Matter} {Coupling} in a
  {Cavity}},\ }\href {https://doi.org/10.1103/PhysRevLett.121.253001}
  {\bibfield  {journal} {\bibinfo  {journal} {Phys. Rev. Lett.}\ }\textbf
  {\bibinfo {volume} {121}},\ \bibinfo {pages} {253001} (\bibinfo {year}
  {2018}{\natexlab{b}})}\BibitemShut {NoStop}%
\bibitem [{\citenamefont {Feist}\ \emph {et~al.}(2018)\citenamefont {Feist},
  \citenamefont {Galego},\ and\ \citenamefont
  {Garcia-Vidal}}]{feist_polaritonic_2018}%
  \BibitemOpen
  \bibfield  {author} {\bibinfo {author} {\bibfnamefont {J.}~\bibnamefont
  {Feist}}, \bibinfo {author} {\bibfnamefont {J.}~\bibnamefont {Galego}},\ and\
  \bibinfo {author} {\bibfnamefont {F.~J.}\ \bibnamefont {Garcia-Vidal}},\
  }\bibfield  {title} {\bibinfo {title} {Polaritonic {Chemistry} with {Organic}
  {Molecules}},\ }\href {https://doi.org/10.1021/acsphotonics.7b00680}
  {\bibfield  {journal} {\bibinfo  {journal} {ACS Photonics}\ }\textbf
  {\bibinfo {volume} {5}},\ \bibinfo {pages} {205} (\bibinfo {year}
  {2018})}\BibitemShut {NoStop}%
\bibitem [{\citenamefont {Ulusoy}\ and\ \citenamefont
  {Vendrell}(2020)}]{ulusoy_dynamics_2020}%
  \BibitemOpen
  \bibfield  {author} {\bibinfo {author} {\bibfnamefont {I.~S.}\ \bibnamefont
  {Ulusoy}}\ and\ \bibinfo {author} {\bibfnamefont {O.}~\bibnamefont
  {Vendrell}},\ }\bibfield  {title} {\bibinfo {title} {Dynamics and
  spectroscopy of molecular ensembles in a lossy microcavity},\ }\href
  {https://doi.org/10.1063/5.0011556} {\bibfield  {journal} {\bibinfo
  {journal} {J. Chem. Phys.}\ }\textbf {\bibinfo {volume} {153}},\ \bibinfo
  {pages} {044108} (\bibinfo {year} {2020})}\BibitemShut {NoStop}%
\bibitem [{\citenamefont {Felicetti}\ \emph {et~al.}(2020)\citenamefont
  {Felicetti}, \citenamefont {Fregoni}, \citenamefont {Schnappinger},
  \citenamefont {Reiter}, \citenamefont {De~Vivie-Riedle},\ and\ \citenamefont
  {Feist}}]{felicetti_photoprotecting_2020}%
  \BibitemOpen
  \bibfield  {author} {\bibinfo {author} {\bibfnamefont {S.}~\bibnamefont
  {Felicetti}}, \bibinfo {author} {\bibfnamefont {J.}~\bibnamefont {Fregoni}},
  \bibinfo {author} {\bibfnamefont {T.}~\bibnamefont {Schnappinger}}, \bibinfo
  {author} {\bibfnamefont {S.}~\bibnamefont {Reiter}}, \bibinfo {author}
  {\bibfnamefont {R.}~\bibnamefont {De~Vivie-Riedle}},\ and\ \bibinfo {author}
  {\bibfnamefont {J.}~\bibnamefont {Feist}},\ }\bibfield  {title} {\bibinfo
  {title} {Photoprotecting {Uracil} by {Coupling} with {Lossy}
  {Nanocavities}},\ }\href {https://doi.org/10.1021/acs.jpclett.0c02236}
  {\bibfield  {journal} {\bibinfo  {journal} {J. Phys. Chem. Lett}\ }\textbf
  {\bibinfo {volume} {11}},\ \bibinfo {pages} {8810} (\bibinfo {year}
  {2020})}\BibitemShut {NoStop}%
\bibitem [{\citenamefont {F{\'a}bri}\ \emph {et~al.}(2024)\citenamefont
  {F{\'a}bri}, \citenamefont {Cs{\'a}sz{\'a}r}, \citenamefont {Hal{\'a}sz},
  \citenamefont {Cederbaum},\ and\ \citenamefont
  {Vib{\'o}k}}]{fabri_coupling_2024}%
  \BibitemOpen
  \bibfield  {author} {\bibinfo {author} {\bibfnamefont {C.}~\bibnamefont
  {F{\'a}bri}}, \bibinfo {author} {\bibfnamefont {A.~G.}\ \bibnamefont
  {Cs{\'a}sz{\'a}r}}, \bibinfo {author} {\bibfnamefont {G.~J.}\ \bibnamefont
  {Hal{\'a}sz}}, \bibinfo {author} {\bibfnamefont {L.~S.}\ \bibnamefont
  {Cederbaum}},\ and\ \bibinfo {author} {\bibfnamefont {{\'A}.}~\bibnamefont
  {Vib{\'o}k}},\ }\bibfield  {title} {\bibinfo {title} {Coupling polyatomic
  molecules to lossy nanocavities: Lindblad vs schr{\"o}dinger description},\
  }\href@noop {} {\bibfield  {journal} {\bibinfo  {journal} {J. Chem. Phys.}\
  }\textbf {\bibinfo {volume} {160}} (\bibinfo {year} {2024})}\BibitemShut
  {NoStop}%
\bibitem [{\citenamefont {Bhuyan}\ \emph {et~al.}(2024)\citenamefont {Bhuyan},
  \citenamefont {Lednev}, \citenamefont {Feist},\ and\ \citenamefont
  {B{\"o}rjesson}}]{bhuyan2024effect}%
  \BibitemOpen
  \bibfield  {author} {\bibinfo {author} {\bibfnamefont {R.}~\bibnamefont
  {Bhuyan}}, \bibinfo {author} {\bibfnamefont {M.}~\bibnamefont {Lednev}},
  \bibinfo {author} {\bibfnamefont {J.}~\bibnamefont {Feist}},\ and\ \bibinfo
  {author} {\bibfnamefont {K.}~\bibnamefont {B{\"o}rjesson}},\ }\bibfield
  {title} {\bibinfo {title} {The effect of the relative size of the exciton
  reservoir on polariton photophysics},\ }\href@noop {} {\bibfield  {journal}
  {\bibinfo  {journal} {Advanced Optical Materials}\ }\textbf {\bibinfo
  {volume} {12}},\ \bibinfo {pages} {2301383} (\bibinfo {year}
  {2024})}\BibitemShut {NoStop}%
\bibitem [{\citenamefont {Sandik}\ \emph {et~al.}(2024)\citenamefont {Sandik},
  \citenamefont {Feist}, \citenamefont {Garc{\'\i}a-Vidal},\ and\ \citenamefont
  {Schwartz}}]{sandik2024cavity}%
  \BibitemOpen
  \bibfield  {author} {\bibinfo {author} {\bibfnamefont {G.}~\bibnamefont
  {Sandik}}, \bibinfo {author} {\bibfnamefont {J.}~\bibnamefont {Feist}},
  \bibinfo {author} {\bibfnamefont {F.~J.}\ \bibnamefont {Garc{\'\i}a-Vidal}},\
  and\ \bibinfo {author} {\bibfnamefont {T.}~\bibnamefont {Schwartz}},\
  }\bibfield  {title} {\bibinfo {title} {Cavity-enhanced energy transport in
  molecular systems},\ }\href@noop {} {\bibfield  {journal} {\bibinfo
  {journal} {Nature Materials}\ ,\ \bibinfo {pages} {1}} (\bibinfo {year}
  {2024})}\BibitemShut {NoStop}%
\end{thebibliography}

%

\end{document}